\documentclass[11pt,preprint]{aastex}
\usepackage{mathrsfs}
\usepackage{amsmath}
\usepackage{empheq}
\usepackage{bm}

\def\bhm{M_{\bullet}}

\def\er{L_{\rm Bol}/L_{\rm Edd}}
\def\ergs{\rm erg~s^{-1}}
\def\fblr{f_{_{\rm BLR}}}
\def\kms{\rm km~s^{-1}}

\def\mathdotM{\dot{\mathscr{M}}}
\def\dotmfactor{\left[1+\ln\left(\mathdotM/\mathdotM_c\right)\right]}

\def\Lhb{L_{_{\rm H\beta}}}

\def\pp{\prime\prime}

\def\prblr{\langle\rblr\rangle}
\def\prblr{R_{_{{\rm H\beta},R-L}}} 
\def\rblr{R_{_{\rm H\beta}}}
\def\rblrl{\rblr-L_{5100}}
\def\rrhb{r_{_{\rm H\beta}}}

\def\Rg{R_{\rm g}}
\def\rhb{R_{_{\rm H\beta}}}

\def\sunm{M_{\odot}}
\def\tauhb{\tau_{_{\rm H\beta}}}

\def\Ythin{Y}
\def\Ysat{Y_{\rm sat}}

\def\feii{Fe {\sc ii}}
\def\mgii{Mg~{\sc ii}}
\def\oiii{[O~{\sc iii}]}

\begin{document}

\title{Supermassive Black Holes with High Accretion Rates in Active Galactic Nuclei. \\
IV. H$\beta$ Time Lags and Implications for Super-Eddington Accretion}

\author
{Pu Du\altaffilmark{1}, 
Chen Hu\altaffilmark{1},
Kai-Xing Lu\altaffilmark{2,1}, 
Ying-Ke Huang\altaffilmark{1}, 
Cheng Cheng\altaffilmark{3},
Jie Qiu\altaffilmark{1},
Yan-Rong Li\altaffilmark{1},
Yang-Wei Zhang\altaffilmark{6},
Xu-Liang Fan\altaffilmark{6},
Jin-Ming Bai\altaffilmark{6}, 
Wei-Hao Bian\altaffilmark{8},
Ye-Fei Yuan\altaffilmark{9},\\
Shai Kaspi\altaffilmark{7},
Luis C. Ho\altaffilmark{4,5},
Hagai Netzer\altaffilmark{7} and
Jian-Min Wang\altaffilmark{1,3,*}\\
(SEAMBH collaboration)}

\altaffiltext{1}
{Key Laboratory for Particle Astrophysics, Institute of High Energy Physics,
Chinese Academy of Sciences, 19B Yuquan Road, Beijing 100049, China}

\altaffiltext{2}
{Astronomy Department, Beijing Normal University, Beijing 100875, China}

\altaffiltext{3}
{National Astronomical Observatories of China, Chinese Academy of Sciences,
 20A Datun Road, Beijing 100020, China}

\altaffiltext{4}
{Kavli Institute for Astronomy and Astrophysics, Peking University, Beijing 100871, China} 

\altaffiltext{5}
{Department of Astronomy, School of Physics, Peking University, Beijing 100871, China} 

\altaffiltext{6}{Yunnan Observatories, Chinese Academy of Sciences, Kunming 650011, China}

\altaffiltext{7}{Wise Observatory, School of Physics and Astronomy, Tel Aviv University, 
Tel Aviv 69978, Israel}

\altaffiltext{8}{Physics Department, Nanjing Normal University, Nanjing 210097, China}

\altaffiltext{9}{Department of Astronomy, University of Science and Technology of China, Hefei 
230026, China}

\altaffiltext{*}{Corresponding author}

\begin{abstract}
We have completed two years of photometric and spectroscopic monitoring of a large number of 
active galactic nuclei (AGNs) with very high accretion rates. In this paper, we report on the 
result of the second phase of the campaign, during 2013--2014, and the measurements of five 
new H$\beta$ time lags out of eight monitored AGNs. All five objects were identified as 
super-Eddington accreting massive black holes (SEAMBHs). The highest measured 
accretion rates for the objects in this campaign are $\mathdotM\gtrsim 200$, where 
$\mathdotM= \dot{M}_{\bullet}/L_{\rm Edd}c^{-2}$, $\dot{M}_{\bullet}$ is the mass accretion 
rates, $L_{\rm Edd}$ is the Eddington luminosity and $c$ is the speed of light. 
We find that the H$\beta$ time lags in SEAMBHs are significantly shorter than those 
measured in sub-Eddington AGNs, and the deviations increase with increasing accretion rates. 
Thus, the relationship between broad-line region size ($\rblr$) and optical 
luminosity at 5100\AA, $\rblrl$, requires accretion rate as an 
additional parameter. We propose that much of the effect may be due to the strong anisotropy 
of the emitted slim-disk radiation. Scaling $\rblr$ 
by the gravitational radius of the black hole, we define a new radius-mass parameter ($\Ythin$) 
and show that it saturates at a critical accretion rate of $\mathdotM_c=6\sim 30$,
indicating a transition from thin to slim accretion disk and a saturated luminosity of 
the slim disks. The parameter $\Ythin$ is a very useful probe for understanding the various
types of accretion onto massive black holes. We briefly comment on
implications to the general population of super-Eddington AGNs in the universe and applications 
to cosmology.
\end{abstract}


\section{Introduction}
Reverberation mapping (RM) experiments, which measure the delayed response of the broad 
emission line gas to the ionizing continuum in active galactic nuclei (AGNs), are the best way 
to map the gas distribution and derive several fundamental properties such as its emissivity-weighted
radius and the black hole (BH) mass. The method was suggested by Bahcall et al. (1972) and its 
theoretical foundation explained in detailed in Blandford \& McKee (1982). Numerous 
RM experiments, since the late 1980s (e.g., Clavel et al. 1991; Peterson et al. 1991, 1993; Maoz 
et al. 1991; Wanders et al. 1993; Dietrich et al. 1993; Kaspi et al. 2000; Denney et al. 2010;
Bentz et al. 2009; Grier et al. 2012; Du et al. 2014; Wang et al. 2014a; Barth et al. 2015), have 
succeeded in mapping the emissivity distribution of H$\beta$ and several other lines in the 
broad-line region (BLR) in more than 40 AGNs. BH virial mass measurements based on the method 
was shown to be consistent with the $\bhm-\sigma_*$ relationship,  or with measurements based 
on stellar dynamics in local objects, where $\sigma_*$ is the stellar velocity dispersion in the 
host galaxy (Ferrarese et al. 2001; Onken et al. 2004; see Kormendy \& Ho 2013 for an extensive 
review). This method is an efficient and routine way to estimate BH mass at basically all redshift, 
at distances that are well beyond the resolving power of all ground-based telescopes.
The RM experiments have been extensively discussed and reviewed in the literature 
(e.g., Peterson 1993; Kaspi et al. 2000 for earlier works and Shen et al. 2015 and De Rosa et al. 
2015 for more recent results). 

Most RM experiments, so far, have focused on the time lag of the broad H$\beta$ line ($\tauhb$ in 
the rest frame) relative to the AGN continuum luminosity ($\lambda L_{\lambda}$) at rest-frame 
wavelength of 5100\AA\ (hereafter $L_{5100}$). Results for 41 such measurements, based on AGN 
luminosity that is corrected for host galaxy contamination, are summarized in Bentz et al. (2013). 
They lead to a simple, highly significant correlation of the form,
\begin{equation}
\rblr\approx \alpha~l_{44}^{\beta}~{\rm ltd} \,\, ,
\end{equation}
where $\rblr=c\tauhb$ is the emissivity-weighted radius of the BLR and 
$l_{44}=L_{5100}/10^{44}\ergs$. We refer to this type of relationship as the $\rblrl$ 
relationship. The constants $\alpha$ and $\beta$ differ slightly from one study to the next, 
depending on the number of sources and their exact luminosity range. For the Bentz et 
al. (2013) work, $\alpha=33.65$ and $\beta=0.533$. Scaling relationships based on other 
emission lines have been used to estimate BH mass in high-redshift AGNs. These are either 
based on the \mgii$\lambda 2798$ line, which is scaled to the H$\beta$ line (e.g., McLure 
\& Dunlop 2004; Vestergaard \& Peterson 2006; Shen et al. 2011; Trakhtenbrot \& Netzer 2012)
or the C {\sc iv}$\lambda 1549$ line, for which few direct lag measurements are available (e.g., 
Kaspi et al. 2007). So far, there is not enough  information about the dependence of the $\rblrl$ 
relationship on accretion rate or Eddington ratio $\er$, where $L_{\rm Bol}$ is the 
bolometric luminosity, $L_{\rm Edd}=1.5 \times 10^{38} (\bhm/\sunm)\ergs$ is the Eddington 
luminosity for a solar composition gas, and $\bhm$  the BH mass.

We are conducting a large RM monitoring campaign targeting high-accretion rate AGNs. The 
aims are to understand better the physical mechanism powering these sources, the dependence 
of the $\rblrl$ relationship on accretion rate, and the possibility to use such objects to 
infer cosmological distances especially at high $z$  (Wang et al. 2013).
We coin these sources ``super-Eddington accreting massive black holes'' (SEAMBHs).
Earlier attempts in this direction, based on monitoring narrow-line Seyfert 1 galaxies (NLS1s), 
failed mostly because of the small variability amplitude of the selected sources (Giannuzzo \&
Stirpe 1996; Giannuzzo et al. 1998; Shemmer \& 
Netzer 2000; Klimek et al. 2004). Bentz et al. (2010) summarized results for a few NLS1s, 
and found relatively small $\er$ for this class of objects and time lags that are not 
significantly different from the ones given by Equation (1). 
Our monitoring campaign started in 2012 and its first two phases are already completed.
The $\sim 20$ targets observed so far are listed in Table 
1. Results from the first phase have been published in Du et al. (2014; hereafter Paper I) 
and Wang et al. (2014a; hereafter Paper II). We also studied the time-dependent variability 
of the strong \feii\, lines in nine of the sources, and the results are reported in Hu et al. 
(2015, hereafter Paper III). We characterize SEAMBHs by their 
dimensionless accretion rate,
$\mathdotM=\dot{M}_{\bullet}/L_{\rm Edd}c^{-2}$,
where $\dot{M}_{\bullet}$ is the accretion rate. Typical values of this parameter range 
from a few to $\sim 100$ for the objects in the first phase of the project. Such high accretion rates
are characteristics of slim accretion disks (Abramowicz et al. 1988) that are thought to power
these objects (Szuszkiewicz et al. 1996; Wang \& Zhou 1999; Wang et al. 1999; Mineshige et al. 2000). 
Paper II shows that such objects may eventually become new standard candles for cosmology. 

This paper reports the results of the second year of our RM campaign. Target selection, 
observation details and data reduction are described in \S2. H$\beta$ lags, BH 
mass and accretion rates are provided in \S3. The new H$\beta$ lags are discussed in \S4, 
showing that for a given luminosity, the H$\beta$ lag gets shorter with increasing $\mathdotM$. 
In \S5 we briefly discuss the implications of the new findings to the understanding of BLR 
physics and geometry, and to accretion physics. \S6 gives a brief 
summary of the paper. Throughout this work we assume a standard cosmology with 
$H_0=67~{\rm km~s^{-1}~Mpc^{-1}}$, $\Omega_{\Lambda}=0.68$ and $\Omega_{\rm M}=0.32$ (Ade et 
al. 2014).

\section{Observations and data reduction}
\subsection{Target Selection}
Unlike the first phase described in Papers I and II, in which most objects are 
NLS1s, for the second phase we selected targets from 
the quasar sample of the Sloan Digital Sky Survey Data Release 7 (SDSS DR7) using  
the pipeline employed in Hu et al. (2008a). The objects have similar spectroscopic 
characteristics to the NLS1s from Papers I and II, in particular:
1) strong optical \feii\ lines, 
2) narrow ($\lesssim 2000\kms$) H$\beta$ lines,
3) weak \oiii\ lines (Osterbrock \& Pogge 1987; Boroson \& Green 1992).
The objects selected for the first year study are NLS1s with extremely steep $2-10$ 
keV continuum. For the second year sample reported here, we do not have X-ray data 
and use, instead, the dimensionless accretion rate $\mathdotM$,  which can be estimated 
through the physics of thin accretion disks as formulated by Shakura \& Sunyaev (1973, 
hereafter SS73). In such systems, the accretion rate can be directly calculated from 
the part of the spectrum where $L_{\nu} \propto \nu^{1/3}$,
regardless of the value of the BH spin (e.g., Collin et al. 2002; Davis \& Laor 
2011 and references therein). The only significant uncertainty in this estimate is the 
disk inclination to the line of sight, $i$ (see more details in Paper II). We take 
$\cos i=0.75$ in this series of papers\footnote{$\cos i=0.75$ represents 
a mean disk inclination for a type 1 AGN with a torus covering factor of about 0.5, assuming 
the torus axis is co-aligned with the disk axis.}. 
The standard thin disk equations give (see Paper II),
\begin{equation}
\mathdotM=20.1\left(\frac{l_{44}}{\cos i}\right)^{3/2}m_7^{-2},
\label{eq:SS_2}
\end{equation}
where $m_7=\bhm/10^7\sunm$. For thin accretion disks, we have $\er=\eta \mathdotM$, where 
$L_{\rm Bol}$ is the disk bolometric luminosity, and $\eta$ is the mass-to-radiation 
conversion efficiency which depends on the BH spin. 

When comparing thin to slim accretion disks it is important to note that in both systems,
the observed 5100\AA\ emission comes from large 
disk radii and thus is less influenced by the radial motion of the accretion flow
compared with the regions closer in that emit the shorter wavelength photons. At these
large radii, Keplerian rotation dominates, radiation cooling locally balances the release of 
gravitational energy through viscosity, and all effects arising from radial advection and 
the black hole spin can be neglected. 
As direct integration of the disk SED is not practical in almost all cases, due to 
the lack of far-UV observations, measuring $\er$ directly is not possible, 
and Equation (2) is the best way to estimate 
$\mathdotM$, which is directly related to the SS73 accretion disk model.

To estimate the BH mass, we followed the standard approach and assume that the BLR gas is moving 
in Keplerian orbits and the rest frame time lags ($\tauhb$) provide reliable
estimates of $\rblr$. This gives,
\begin{equation}
\bhm=\fblr \frac{\rblr V_{\rm FWHM}^2}{G}=1.95\times 10^6~\fblr V_{3}^2\tau_{10}~\sunm,
\end{equation}
where $G$ is the gravitational constant, $V_{3}=V_{\rm FWHM}/10^3\kms$ is the 
full-width-half-maximum (FWHM) of the H$\beta$ line profile in units of 
$10^3\kms$ and $\tau_{10}=\tauhb/10$days. 
The factor $\fblr$ is calibrated from the known $\bhm-\sigma$ relation (e.g., Onken et al. 
2004; Ho \& Kim 2014). This is still a matter of some debate and the quoted uncertainties are 
large. The study of Ho \& Kim (2014) shows that for AGNs in host galaxies with pseudo-bulges, 
$\fblr$ is smaller than in AGNs hosted by classical bulges or ellipticals. However, $\fblr$ gets 
larger for AGNs with higher Eddington ratios. It is not very clear what is the end result of 
these two opposite trends. For most SEAMBHs in Papers I and II, the host galaxies were observed 
by the Hubble Space Telescope ({\it HST}) and show indications for pseudo-bulges. They also have 
very high $\er$. For the present sample we do not have {\it HST} images but these sources also 
are of high $\er$. Given all this, we use $\fblr=1$, as in Papers I and II, but note the large
uncertainty on this number.

Employing Equations (1) and (3), we estimated $\mathdotM$ from Equation (2) for quasars in the 
SDSS DR7 and selected sources with the highest $\mathdotM$. We discarded radio-loud objects based 
on the reported {\it FIRST} observations. We also chose a redshift range of $z=0.1-0.3$ and 
$L_{5100}\lesssim 10^{44.5}\ergs$ to make sure that the lag, as estimated from 
Equation (1), can be measured in $5-6$ months of monitoring campaign and magnitude 
$r^{\prime}\le 18.0$ to ensure high enough signal-to-noise (S/N). Details of the sources 
are given in Table 1.

\subsection{Photometry and Spectroscopy}
The second year observations described here are essentially identical to those of the first 
year, and readers are referred to Papers I and II for detailed information about the telescope 
and spectrograph. In short, the telescope is a 2.4 m alt-azimuth mounted Ritchey-Chr\'etien 
located in Lijiang, Yunnan Province, and operated by the Yunnan Observatories. We 
use the Yunnan Faint Object Spectrograph and Camera (YFOSC) with a back-illuminated 
2048$\times$4608 pixel CCD covering a field of $10^{\prime}\times 10^{\prime}$.
The differences between the first and 2nd year observations are: 1) we adopted 
a $5^{\pp}$-wide slit, compared with the previously used $2.5^{\pp}$ slit, to minimize the influence 
of atmospheric differential refraction, and used Grism 3 with spectral resolution of 2.9\AA/pixel 
and wavelength coverage of 3800-9000\AA. 2) Instead of the Johnson $V$ filter used in 2012, 
we used SDSS $r^{\prime}$-band filter to avoid the 
potential contamination by emission lines such as H$\beta$ and H$\alpha$. 3) The observations now 
include photometry from the Wise Observatory, 1m telescope in Israel, where we used a Princeton
Instruments CCD camera, an SDSS $r^{\prime}-$band filter and a typical exposure time of 900 sec. 
The reduction of the Wise and Yunnan Observatory photometry data was done in a standard way using 
{\it IRAF} routines. For the Lijiang data, the reduction, and the procedures to measure 5100\AA\ 
and F(H$\beta$), are given in Papers I while for Wise data, 
the flux measurements were done using point-spread function photometry. The light curves were 
produced by comparing the instrumental magnitudes to those of constant stars in the field (see, 
e.g., Netzer et al. 1996, for details). The uncertainties on the photometric measurements include 
the fluctuations due to photon statistics and the scatter in the measurement of the stars used. 
Detailed information of observations, such as, comparison stars etc. is provided in Table 1. We 
also list the critical S/N used as a low threshold to reject poor weather condition observations.

The 5100\AA\ light curves used in the analysis were obtained by inter-calibration of the 5100\AA\ 
spectroscopy, $r^{\prime}-$band Lijiang 2.4-m photometry, and $r^{\prime}-$band Wise Observatory 
photometry. We used the Bayesian algorithm described in Li et al. (2014), who assume that the AGN 
optical continuum variations follow a damped random walk model (Kelly et al. 2009; Li et al. 2013). 
We applied a multiplicative scale factor and an additive flux adjustment to bring the different 
measurements to a common flux scale. The algorithm performs inter-calibration on all the datasets 
simultaneously. This enables us to relax the requirements for the sampling rates and retain the 
highest achievable temporal resolution. We selected the Lijiang spectroscopy as the common scale. 
The best parameters for the inter calibration are determined by a Markov chain Monte Carlo (MCMC) 
implementation. In that way we get the inter-calibrated continuum light curves of photometry (Lijiang 
and Wise data) and spectral continuum (Lijiang data). Finally, the combined light curves 
are obtained by averaging all the points of a same day in the inter-calibrated light curves. All 
the continuum and H$\beta$ light curves for the eight objects are listed in Tables $2-5$, and shown 
in Figure 1. We also calculated the mean and RMS (root mean square) spectra and present them in 
Appendix A where we briefly discuss these data.

\subsection{Host Galaxies}
Most of the objects described in Papers I and II had previous HST images that were used to subtract 
the host galaxy contribution at 5100\AA. Such information is not available for the sources 
reported in this paper. Instead, we used the empirical relation suggested by Shen et al. (2011) 
based on SDSS spectra: $L_{5100}^{\rm host}/L_{5100}^{\rm AGN}=0.8052-1.5502x+0.912x^2-0.1577x^3$, 
for $x<1.053$, where $x=\log \left(L_{5100}^{\rm tot}/10^{44}{\rm erg~s^{-1}}\right)$ and 
$L_{5100}^{\rm tot}$ is the total emissions of AGNs and their host at 5100\AA. For  $x>1.053$ we 
have $L_{5100}^{\rm host}\ll L_{5100}^{\rm AGN}$, and the host contamination can be neglected.
These authors obtained the geometric mean composite spectra for objects binned in $\log L_{5100}$ 
normalized at 3000\AA\, and found spectral flattening toward long wavelengths. The flatten part of 
the spectra depends on host contaminations. This empirical relation was obtained by the difference 
of the flatten part through comparisons among the spectra for quasars in 5100\AA\, luminosity bins. 

The above procedure was developed for the $3^{\pp}$ SDSS fiber noted as $x_{_{3^{\pp}}}$. 
This is considerably smaller than the effective aperture ($5^{\pp}$) used by us. Denote  
$x_{_{5^{\pp}}}$ as the flux measured from our mean spectra
and  $\Delta x=|x_{_{3^{\pp}}}-x_{_{5^{\pp}}}|$. We find that on average 
$\Delta x/x_{_{3^{\pp}}} < 0.3$. We also find that the
differences in $L_{5100}^{\rm host}/L_{5100}^{\rm AGN}$ between the  $3^{\pp}$ fiber and 
our $5^{\pp}$ aperture is less than 10\%. 
The fraction is estimated by comparing $F_{5^{\prime\prime}}$ 
and $F_{3^{\prime\prime}}$ fluxes at 5100\AA\, in our sample, where $F_{5^{\prime\prime}}$ and 
$F_{3^{\prime\prime}}$ are measured from
Lijiang and SDSS observations, respectively. In general, more than 90\% of the light is contained 
within the $3^{\prime\prime}$ fiber for $z\sim 0.2$ quasars.
Thus, the present subtraction of host contamination is quite robust, and 
the added uncertainty due to this 
difference is not larger than the intrinsic uncertainty in Shen et al.'s expression.

\section{H$\beta$ Lags, Black Hole Mass and Accretion Rates}
We used a standard cross-correlation analysis to determine the lags of the H$\beta$ line relative 
to the combined 5100\AA\ continuum. The procedure is quite standard and was described, in 
detail, in Papers I and II, where we also list all the relevant references. The uncertainties on 
the lags are determined through the ``flux randomization/random subset sampling" method (RS/RSS) 
(Peterson et al. 1998, 2004), the cross-correlation centroid distribution (CCCD) and 
cross-correlation peak distribution (CCPD) generated by RS/RSS method (Maoz \& Netzer 1989; Peterson 
et al. 1998, 2004; Denney et al. 2006, 2010 and reference therein), which are shown in Figure 1.
For a successful detection of H$\beta$ lag we require: 1) non-zero lag
from the CCF peak and 2) a maximum correlation coefficient larger than 0.5.

Five of the eight sources show well-determined lags. All these sources are listed in
Table 6. The other three failed our above criteria. Figure 1 
shows that the CCFs of two of them (J094422 and J100055) show two comparable peaks and no significant 
H$\beta$ lags. The third source, J080131, has an unusual combination of line and continuum light curves. 
The continuum light curve shows a well-determined, broad feature followed by two major short "bursts" 
centered at around JD24566000+, $120$ and $140$.
The H$\beta$ light curve shows a clear response to the first continuum dip and
burst feature, but no response to the two shorter duration features. The CCF of the entire campaign 
has a low peak, at $r_{\rm max} \simeq 0.5$, which is consistent with a zero lag. Carrying the 
analysis for the first 70 days only, as shown in the second diagram of J080131 panel in Figure 1, 
we found a very significant peak with a centroid lag of $11.5_{-3.6}^{+8.4}$ days (with a very high
coefficient of $r_{\rm max}=0.81$) in rest-frame. 

Our search of the literature shows that the unusual combination of line and continuum light curves 
observed in J080131 is rare 
[see similar, but not identical, behavior in NGC 7469 (Peterson et al. 2014) 
and in the UV light curve of  NGC\,5548 (De Rosa et al. 2015)]. 
We can think of various scenarios that could cause such an event in a SEAMBH with extremely 
high accretion rate.  For example, the inner part of slim disks can be so thick that
self-shadowing effects lead to strong anisotropy of the emitted radiation.
The signal received and measured by a remote observer can differ substantially from the 
ionizing radiation that reaches the H$\beta$-emitting clouds (see discussions by Wang et al. 2014b).
Given the rarity of such objects, we have no 
way to test this idea and thus do not include this object in the remaining analysis. 

We used Equations 2 and 3 to calculate accretion rates, $\mathdotM$ and BH masses for the five 
sources listed in Table 6. The details of the procedure are given in papers I and II and the
distributions of $\mathdotM$ in the two groups of the previously 
and currently monitored objects are shown in Figure 2. As clearly shown in the diagram, 
our Lijiang-observed sources occupy the high tail of the $\mathdotM$ distribution. This is not 
surprising given the way we selected our targets. We also show in Figure 2 the equivalent width (EW) 
distribution of H$\beta$. On average, our high accretion rate sources have lower mean EW(H$\beta$). 
An anti-correlation between EW(H$\beta$) and $L_{\rm Bol}/L_{\rm Edd}$ has been found
in other samples with less accurate rates, in which BH mass is estimated by the $\rblrl$ relation 
(e.g., Netzer et al. 2004).

In Paper II we classified SEAMBHs as those objects with $\dot{m}=\eta\mathdotM \geq 0.1$.
This is based on the idea that beyond this value, the accretion disk becomes slim and the radiation 
efficiency is reduced due to photon trapping and other effects. Since we cannot observe the entire 
SED, we have no direct way to measure $\er$, and this criterion is used as an approximate tool 
for identifying SEAMBH candidates. To be on the conservative side, we chose the lowest possible 
efficiency, $\eta=0.038$ (retrograde disk with $a=-1$, see Bardeen et al. 1972). Thus SEAMBHs 
are objects with $\mathdotM=2.63$. 
For simplicity, in this paper we use $\mathdotM_{\rm min}=3$ as the required minimum. 
Later on, in \S5, we introduce an empirical way, based on our own mass measurements, to 
define this group more accurately.
 
Table 6 lists all observables, $\bhm$ and $\mathdotM$ for the first (SEAMBH2012) and second 
(SEAMBH2013) year observations. All objects, except for MCG\,$+06-26-012$ (see note
in the table caption), show $\mathdotM>3$, indicating they are SEAMBHs.
In Paper III, more accurate measurements of H$\beta$ lags have been derived by using a 
scheme of simultaneously fitting spectra of host and AGNs. The newly measured lags are the ones
listed table 6. Comparing with the previous measurements in Papers I and II, the updated 
lags are consistent with the previous ones, but the error bars are much smaller. Including 
our new observations, the majority of SEAMBHs with RM-based mass measurements come from our 
Lijiang RM campaign.

\section{ H$\beta$ Time Lags in SEAMBHs}
To understand better the fundamental differences between high and low $\mathdotM$ 
sources, we must use the measured correlation between time lag and source luminosity for 
all RM AGNs. For this, we have to take into account that some of the sources have
been mapped more than once and hence there are various measured values for their luminosity, 
time lag, and FWHM(H$\beta$), for the same BH mass.  There are two possible ways to 
address this issue. The first is to average the BH mass from the individual campaigns, and 
then obtain mean (or median) values for $L_{5100}$, $\Lhb$ and the time lag. This scheme, 
which was used in Kaspi et al (2005), Bentz et al (2009) and other papers, is referred here 
as the ``average scheme''. The second is to treat the independent RM campaigns of a single 
object as different objects (e.g., Bentz et al. 2013). We call this the ``direct scheme''.

The goal of the present paper is to find the basic properties of the central power house, 
presumably an accretion disk, as a function of BH mass and normalized accretion rate. The 
measured H$\beta$ time lag is the tool we use to study these properties. This is changing 
with the ionizing continuum but it does not represent real changes in the gas distribution 
in the BLR during one observing season. 
Since the BH mass does not change on a short time scale, and the time scale to change the 
global accretion rate through the disk is also very long, we must be seeing short time 
scale fluctuations that do not change these fundamental properties but, nevertheless, affect 
the ionization of the BLR. To understand the population properties, we must give equal weights 
to all objects, i.e., use the average scheme approach. This means that the mean luminosity is 
used to derive $\mathdotM$. This can be done during one campaign that lasts a few months, or 
during several campaigns that last a few years. The only exception is, perhaps, when the total 
time exceeds the dynamical time scale of the BLR.  

The direct scheme can be useful to derive other properties, such as the ionization distribution 
across a ``typical BLR''. While this is not the goal of the present work, we have calculated 
all the fundamental correlations discussed below using this approach, too, and show them in
Appendix B. We note that despite the fundamentally 
different approaches, which result in a different BH mass distribution (since several
masses are counted more than once), the results are not very different.

\subsection{Continuum luminosity $ vs. $ BLR Size}
Table 7 lists the objects summarized by Bentz et al. (2013), plus a few newly mapped sources. 
Since some of the procedures used in earlier campaigns differ in several aspects, we decided to 
apply the same uniform procedure for all the sources. For objects with multiple RM
measurements, our procedure for obtaining the average values is as follows: We first calculate 
$\bhm$ for each of the individual campaigns and then average all these values, using weighted
standard deviations,  to get a mean $\bhm$ and its error. This value is used to obtain individual 
values of $\mathdotM$ for the various campaigns, using $L_{5100}$ from this campaign. 
The mean $\mathdotM$ for the object is the average of the individual $\mathdotM$, and its 
errors include the standard deviation. For the mean luminosity and $\tauhb$, we used logarithmic 
averages.

Regarding BH mass calculations in individual campaigns, this is more problematic 
since some works used the RMS spectrum (e.g., Peterson et al. 2004; Bentz et al. 2009b; Denney et 
al. 2010; Grier et al. 2012) while others prefer the use of FWHM(H$\beta$) 
(e.g., Kaspi et al. 2005). Our procedure is based on the FWHM method (see explanation in Papers I 
and II). As shown recently in Woo et al. (2015), the scatter in the scaling parameter ($f_{\rm BLR}$)
derived in this method is very similar to the scatter
in the method based on the RMS spectrum.

We have therefore collected from the literature all the required information about the 
H$\beta$ line and carried out our own mass measurements for the sources where earlier mass 
measurements were based on the RMS method. In most of the cases, the agreement between the 
older RMS method and the value obtained here is within the uncertainty on the BH mass. In a few 
cases (five objects: NGC 4051, NGC 4151, PG 0804+761, PG 1613+658 and PG 2130+099) the deviation 
is larger which we interpret as unrealistically
small uncertainty on the mass. The deviation is generally smaller than $\sim 0.5$
dex except for NGC 4051 where the deviation is $\sim 0.7$dex.
Some more details for FWHM measurements of a few objects are given in the caption 
of Table 7. We list all the measured values and their averages in this table\footnote{Following Bentz 
et al. (2013), we included in the uncertainties of the 5100\AA\, luminosities also the uncertainties
due to distances.}. 

All correlations shown in this paper were calculated with the FITEXY method in the version 
adopted by Tremaine et al. (2002), where scatter is allowed for by increasing the uncertainties in 
small steps until $\chi^2$ is about unity (this is typical for many of our correlations). We also 
used the BCES method (Akristas \& Bershady 1996) but prefer not to use its results since it is known 
to give unreliable results in samples containing a few outliers (there are a few objects with quite 
large uncertainties of $\mathdotM$). 

Figure 3{\it a} shows H$\beta$ lags versus $L_{5100}$ for all the observations of mapped AGNs 
(i.e., multiple measurements for each object when available). Our objects are marked with blue 
(first year, data reported in Papers I and II) and red (second year, this paper). 
Using the data in Table 7 we get:
\begin{subequations}
\begin{empheq}[left={\log \left(\rhb/{\rm ltd}\right)=\empheqlbrace}]{align}
&(1.45\pm 0.03)+(0.50\pm 0.03)\log l_{44}&({\rm entire~sample}),\\
&(1.54\pm0.03)+(0.53\pm0.03)\log l_{44}&(\mathdotM<3),\\
&(1.30\pm 0.05)+(0.54\pm 0.06)\log l_{44}&(\mathdotM\ge3),
\end{empheq}
\end{subequations}
with intrinsic scatters $\sigma_{\rm in}=(0.21,0.15,0.24)$ for (4a, 4b, 4c), respectively.
The regression for the $\mathdotM<3$ sample is almost identical to Bentz et al. (2013). Equations 
(4b) and (4c) show  different intercepts for the $\mathdotM<3$ and the $\mathdotM\ge3$ samples. 
Most of the SEAMBHs from our campaign are located below the $\rblrl$ relation, 
increasing the scatter in the relationships.
For more extreme SEAMBHs with $\mathdotM>10$, the regression
shows $\log (\rhb/{\rm ltd})=(1.18\pm0.07)+(0.48\pm0.10)\log l_{44}$, indicating they deviate from 
the $\mathdotM<3$ group more than the $\mathdotM\ge3$ group. It is clear that 
the SEAMBH sources increase the scatter considerably, especially over the limited luminosity range 
occupied by the new sources. 

In the following analysis, we use the averaged lags and the averaged mass as listed in Table 7 
(note that unlike in paper-II, here we include in the analysis the few radio-loud AGNs with measured 
time lags and BH mass). In Figure 3{\it c}, we divided the population into two sub-groups, those 
with  $\mathdotM\ge 3$ (11 from earlier studies and 13 from our study) and those with $\mathdotM<3$. 
The diagram emphasizes that much of the intrinsic scatter is caused by a systematic deviation of the
large $\mathdotM$ sources towards shorter lags. The regression calculations give 
\begin{subequations}
\begin{empheq}[left={\log \left(\rhb/{\rm ltd}\right)=\empheqlbrace}]{align}
&(1.45\pm 0.04)+(0.50\pm 0.03)\log l_{44}&({\rm entire~sample}),\\
&(1.55\pm0.04)+(0.53\pm0.04)\log l_{44}&(\mathdotM<3),\\
&(1.32\pm 0.06)+(0.53\pm 0.06)\log l_{44}&(\mathdotM\ge3),
\end{empheq}
\end{subequations}
with intrinsic scatters of $\sigma_{\rm in}=(0.21,0.16,0.22)$ for (5a, 5b, 5c), respectively.
The slope of the correlation for the SEAMBH sample, those with $\mathdotM\ge3$, is comparable 
to that of sub-Eddington AGNs, but the normalisation is significantly different.
Clearly, the H$\beta$ region of SEAMBHs is smaller than 
sub-Eddington AGNs (see the different intercepts in Equations 5b and 5c).

A more complete approach is to analyze the dependence of $\tauhb$ on $L_{5100}$ and $\mathdotM$ 
together, however, several reasons prevent us from performing such an analysis. 
First, the current sample is still small; and second the luminosity range of SEAMBHs is very 
narrow. There are only 4 luminous SEAMBHs (PG 0026+129, PG 1211+143,  PG 1226+023, PG 1700+518)
identified from the previous campaigns, but their $\mathdotM=(4.5, 6.9, 5.0, 12.0)$, respectively,
are smaller than the average of this group. Thus, the search for a new relationship of the type of
$\rhb=\rhb(L_{5100},\mathdotM)$ will have to wait for future observations of luminous, high accretion 
rate sources.

\subsection{H$\beta$ luminosity $vs.$ BLR Size}
Figure 3{\it c} shows the new $\rhb-\Lhb$ relationship. Such relationships 
have been used in the past to supplement the continuum-based relationships and to correlate 
the BLR size more closely with the ionizing continuum radiation (see e.g., Kaspi et al. 2005).
The regression analysis gives,
\begin{subequations}
\begin{empheq}[left={\log \left(\rhb/{\rm ltd}\right)=\empheqlbrace}]{align}
&(1.30\pm 0.03)+(0.51\pm0.03)\log L_{\rm H\beta,42}&({\rm entire~ sample}),\\
&(1.36\pm0.04)+(0.52\pm0.04)\log L_{\rm H\beta,42}&(\mathdotM<3),\\
&(1.21\pm0.06)+(0.53\pm0.06)\log L_{\rm H\beta,42}&(\mathdotM\ge3),
\end{empheq}
\end{subequations}
where $L_{\rm H\beta, 42}=\Lhb/10^{42}~{\rm erg~s^{-1}}$, and 
$\sigma_{\rm in}=(0.20,0.18,0.22)$ for (6a,6b,6c), respectively. 
As in the continuum luminosity case, the inferred BLR size for SEAMBHs 
is smaller than the size of the sources with $\mathdotM < 3$
but the differences are smaller in this case.

The $\rhb-\Lhb$ relationship has been examined by Wu et al. (2004) and Kaspi et al. 
(2005). A comparison with the results of these papers shows similar intercepts  
but different slopes, with the present slopes being significantly flatter
than $0.685\pm 0.106$ found in Wu et al. (2004) and $0.694\pm 0.064$  found by Kaspi 
et al. (2005). It is not at all clear that these are significant variations since 
the present sample is much larger and the distribution of sources along the $\Lhb$ 
axis is quite different. The correlations involving 
$L_{5100}$ and $\Lhb$ (Figure 3 and Figure 9 in
Appendix B) show a very similar scatter but, for a given H$\beta$ luminosity, SEAMBHs 
have shorter lags compared with low-accretion rate AGNs. Such differences may be 
related to the properties of slim accretion disks (Wang et al. 2014b), or other types 
of anisotropies (e.g., Netzer 1987; O'Brien et al. 1994; Goad \& Wanders 1996; 
Ferland et al. 2009). 

\subsection{$\mathdotM-$dependent BLR Size}

To test the dependence of the BLR size on accretion rates, we define a new parameter,  
$\Delta \rhb=\log\left(\rhb/\prblr\right)$ 
that specifies the deviations of individual objects from the $\rblrl/\Lhb$ relationship of the 
sub-sample of $\mathdotM<3.0$ sources (i.e., $\prblr$ as given by Equations 4b, 5b and 6b). 
The scatter of $\Delta \rhb$ is calculated by  
$\sigma=\left[\sum_i\left(\Delta R_{{\rm H\beta}, i}-\langle\Delta R_{{\rm H\beta}}\rangle\right)^2/N\right]^{1/2}$, 
where $N$ is the number of objects and $\langle \Delta R_{{\rm H\beta}}\rangle$ is the averaged 
value. Figure 3 provides the value of $\sigma$ for comparison.

Figure 4 shows plots of $\Delta \rhb$ versus $\mathdotM$, and $\Delta \rhb$ distributions
for the $\mathdotM\ge3$ and $\mathdotM<3$ samples. A Kolmogorov-Smirnov (KS) test comparing 
the two shows that the probability of the same parent distributions is $p_{_{\rm KS}}=0.0054$ 
for the  $\rblrl$ case and $p_{_{\rm KS}}=0.014$ for $\rhb-\Lhb$ case. This provides a strong 
indication that the main cause of  deviation from the old $\rblrl$ relationship (or the $\rblrl$
relationship for the sub-Eddington AGNs) is the extreme accretion rate. Thus, a single $\rblrl$
relationship for all AGNs is a poor approximation for a more complex situation where both the 
luminosity and the accretion rate determine this relationship. From the regression, we get
the
dependence of the deviations of $\rhb$ from the $\rblrl$ relation in Figure 4a 
\begin{equation}
\Delta \rhb=(0.36\pm0.13)-(0.46\pm0.10)\log \mathdotM ~~~ ({\rm for ~\mathdotM\ge3} ),
\end{equation}
with $\sigma_{\rm in}=0.07$. From the $\rblr-L_{\rm H\beta}$ relation in Figure 4c, we get
\begin{equation}
\Delta \rhb=(0.33\pm0.15)-(0.38\pm0.10)\log \mathdotM ~~~({\rm for~\mathdotM\ge3}),
\end{equation}
with $\sigma_{\rm in}=0.13$. We have tested the above correlations also for $\mathdotM<3$. 
The FITEXY regressions give slopes of around zero with very large uncertainties, 
$\Delta \rhb\propto \mathdotM^{-0.085\pm 0.051}$ and 
$\Delta \rhb\propto \mathdotM^{-0.059\pm0.054}$ for Figure 4a and 4c, respectively, implying 
that $\Delta\rhb$ does not correlate with $\mathdotM$ for $\mathdotM<3$ group. All this confirms 
that $\mathdotM$ is an additional parameter controlling the $\rblrl$ relation in AGNs with 
high accretion rates.

To summarize, the new SEAMBHs observed in 2012 and 2013 significantly increase
the scatter of the $\rblrl$ relation. We find that SEAMBHs have significantly shorter lags 
than those of sub-Eddington AGNs with similar 
5100\AA\, luminosity, and the shortened lags increase with the dimensionless accretion rate.
Given this, we recommend to use Equation (4b) or (5b) when trying to estimate the BLR 
size in sub-Eddington AGNs. We suggest that the dependence of $\rhb$ on the dimensionless 
accretion rate could be a consequence of the anisotropic radiation of slim disks. We come back 
to this issue in the following section.

\section{Discussions}
The new results presented here illustrate that the small scatter in the earlier $\rblrl$ 
relationships (Kaspi et al. 2005; Bentz et al. 2013 and references therein),
only applies to low accretion rate AGNs, which, according to our definition, are sources 
with $\mathdotM<3$. The addition of high accretion rate sources, those referred to as SEAMBHs, 
changes this picture. The scatter increases and there is a clear, 
statistically significant tendency for AGNs with higher accretion rates to show smaller 
$\rblr$. Moreover, the deviation increases 
with $\mathdotM$. This holds for both $L_{5100}$ and $\Lhb$ although the 
deviations are larger when using the 5100\AA\ luminosity. The deviations are most noticeable  
over the luminosity range where most SEAMBHs are concentrated, between $10^{43}$ and 
$few\times 10^{44}\ergs$. Below we explore the physical implications of the new results
focusing on the earlier suggestion (Paper II) that SEAMBHs are powered
by slim accretion disks with properties that are considerably different from those of  
thin disks.

\subsection{$\mathdotM$ and the Origin of the Shorter Time Lag}
The deviations from the earlier $\rblrl$ relationship raise several interesting possibilities 
that can explain the new results. The first possibility is that the ionizing 
flux reaching the BLR, which determines its RM-measured size, is different from the flux inferred 
by a remote observer due to an unusual SED and/or the anisotropy of the radiation emitted from the 
disk. This possibility has been mentioned in the past with regards to the covering factor and the 
level of ionization of the BLR. Netzer (1987) presented a detailed model where the combination of 
an angle-dependent SED of geometrically thin accretion disks (see also Czerny \& Elvis 1987; Li et 
al. 2010 for geometrically thick parts of the disks), with an isotropic X-ray source, results 
in a weaker ionizing continuum along the plane 
of the disk compared with the pole-on direction. This will result in a complicated  angle-dependent 
level of ionization. The recent work by Wang et al. (2014b) suggests that large SED variations, 
especially in the ionizing luminosity as a function of the polar angle, will be most noticeable 
in slim accretion disks where the anisotropy of the disk is more extreme because of self-shadowing 
effects. Such a geometry would lead to a smaller emissivity-weighted BLR radius at large angles 
relative to the polar axis. The second possibility is that much of the effect is due to the change 
of the bolometric luminosity with polar angle due to the reasons discussed above. This will reduce 
the dust sublimation radius that has a large effect on the measured RM size (see below). Obviously, 
the covering factor of the BLR, which determines $\Lhb$, and the 
anisotropy of the Balmer line emission, could also affect these relationships.

The search for a direct ionizing luminosity indicator  
initiated several attempts to replace $L_{5100}$ by $\Lhb$ in the $\rblrl$ relationship because 
H$\beta$ responds more directly to the ionizing luminosity (e.g., Peterson 1993). 
Kaspi et al. (2005) show that this 
substitution had no affect on the deduced BLR size. Similar ideas have been proposed by Wang 
and Zhang (2003), Wu et al. (2004), and others. Our new observations hint to the possibility 
that the changes in the $\Lhb$ relationship relative to the relationship
for the $\mathdotM<3$  group are smaller than in the $L_{5100}$ case. Unfortunately,  
the relatively small size of the SEAMBH sample prevents 
us from quantifying these ideas.

\subsection{Accretion Disks and the Maximum Size of the BLR}
\subsubsection{Geometrically thin disks}
Here we explore the possibility that the observed changes in BLR as a function of $\mathdotM$ 
are due to the transition from thin ($\mathdotM<3$) to slim ($\mathdotM\ge3$) disks. To start,
we express the measured  H$\beta$ lags in units of the gravitational radius, 
$\Rg=1.5\times 10^5(\bhm/\sunm)$ cm. Since $\rblr \propto l_{44}^b$ (Equation 1), and 
$l_{44}\propto \mathdotM^{2/3}\bhm^{4/3}$ (Equation 2, recast), we expect 
\begin{equation}
\rrhb=\frac{\rhb}{\Rg}=1.9\times 10^4~\mathdotM^{0.35}m_7^{-0.29}, 
\end{equation}
where we use Equation (5b). Equation (9) concisely mergers the empirical $\rblrl$ relation 
with the basic expectations of geometrically thin disks.

We now defined a new parameter, $\Ythin=m_7^{0.29}\rrhb$, which 
we coin ``the radius-mass parameter", i.e.  
\begin{equation}
\Ythin=1.9\times 10^4 \mathdotM^{0.35}.
\end{equation}
We calculated this combined parameter for all sources in order to test whether 
it depends on accretion rate. 
This test would provide clues about the various modes of accretion and their relationship with
the size of the BLR.

Similar to the definition of $\Delta \rhb$ in Section 4.3,  we define 
$\Delta X=\log \left(X/X_{\rm reg}\right)$, where $X$ is the measured value for individual
sources and $X_{\rm reg}$ is the value obtained for the group from the regression analysis.
In this section we consider two cases,   $X=\rrhb$ and $X=\Ythin$.
The various tests performed on the accretion rate groups are shown in Figure 5 and discussed below. 
 
Our first test for $\rrhb$ is performed on the two accretion rate groups, $\mathdotM<3$ 
and $\mathdotM\ge3$. Figure 5{\it a} shows the $\rrhb-\mathdotM$ relation.  For $\mathdotM<3$ objects,
we find $\rrhb \propto \mathdotM^{0.50\pm 0.08}$ with $\sigma_{\rm in}=0.24$, which is 
steeper than expected (Equation 9). By contrast, $\rrhb\propto \mathdotM^{0.35\pm 0.12}$ 
for the $\mathdotM\ge 3$ group, consistent with Equation (9). 
It is clear that the $\rrhb-\mathdotM$ relation is markedly different for the two groups, 
suggesting that SEAMBHs are physically distinct from sub-Eddington AGNs. $\Delta\rrhb$ and 
its distributions are shown in the lower panels of Figure 5{\it a}. A KS test gives the
probability of $p_{_{\rm KS}}=1.3\times 10^{-3}$ and demonstrates the differences again
for the $\mathdotM\ge3$ and the $\mathdotM<3$ groups. We suspect that the deviations from 
the expected dependences on $\mathdotM^{0.35}$ are due to the dependence
on $m_7$ in our sample that contain a large range of BH mass.

Figure 5{\it b} shows the $\Ythin-\mathdotM$ relation, which has much smaller scatter
than the $\rrhb-\mathdotM$ relation, for both groups.
The regression gives $\Ythin\propto \mathdotM^{0.41\pm0.07}$ for the $\mathdotM<3$ group with 
$\sigma_{\rm in}=0.15$, which, within the errors, agrees well with Equation (10) 
confirming the idea of thin accretion disks in objects with $\mathdotM<3$.
The diagram  clearly shows a change of slope at a transition accretion rate of 
$\mathdotM_c\gtrsim 3$ from a rising to an almost horizontal constant (saturated) value. This 
transition corresponds to a saturated$-Y$ of
$Y_{\rm sat}\sim 3\times 10^4$. This is consistent with the idea that beyond this 
critical rate, which we name $\mathdotM_c$,
the central power-house is no longer a thin accretion disk. Obviously, such a
test could not be performed without the new observational data from SEAMBH2012 and SEAMBH2013.
This is not entirely unexpected since previous works specifies in the introduction to this paper, 
and explain in more detail below, show that the SS73 disk model becomes invalid above a certain
accretion rate (e.g., Laor \& Netzer 1989).

\subsubsection{The Transition Accretion Rate}
Figure 6 shows the determination of the transition accretion rate and the saturated$-Y$.
To determine $\mathdotM_c$ from the data, we first test the dependence of the index $b$ of 
the $\rblrl$ relation on $\mathdotM_c$, and find $b=(0.527\pm0.038,0.528\pm0.030,0.533\pm0.028$) 
for samples of objects with $\mathdotM\le\mathdotM_c=(3,10,20)$. These changes are so small that 
we decided to adopt $b=0.53$ for all the groups. In order to obtain $\mathdotM_c$ and $\Ysat$, we 
use the following functional form\footnote{We also tried to fit the data with a broken power-law, 
$Y=Y_{\rm sat}\mathdotM^{k_1}[1+(\mathdotM/\mathdotM_c)^{k_2}]^{-1}$, where $k_1$, $k_2$, $Y_{\rm sat}$ 
and $\mathdotM_c$ can be obtained from the fit. This did not result in a satisfactory fit. 
The reasons are: 1. Many more objects with low $\mathdotM$ regardless of the exact value of $\mathdotM_c$. 
2. The overall range in $\mathdotM$ for 
low accretion rate sources (more than 3.5 dex) is much larger than for high accretion rate
sources (less than 1.5 dex). }.
\begin{equation}
Y=\left\{\begin{array}{ll}
    \displaystyle{\Ysat\left(\frac{\mathdotM}{\mathdotM_c}\right)^{k_0}}& (\mathdotM<\mathdotM_c),\\
    \Ysat         & (\mathdotM\ge \mathdotM_c),\end{array}
    \right.
\end{equation}  
where $\Ysat$, $k_0$ and $\mathdotM_c$ are to be determined by the data. 
We make use of the Levenberg-Marquardt method to fit the entire sample
with the inclusion of error on $\mathdotM$ through Monte-Carlo simulations (e.g., Press et al. 1992). 
In the fitting, we increase the uncertainties of $Y$ until $\chi^2=1$. This gives
\begin{equation}
\mathdotM_c=13.8_{-8.1}^{+19.6},~~\Ysat=3.3_{-0.5}^{+0.6}\times 10^4,
~~{\rm and}~~k_0=0.27\pm 0.04.
\end{equation} 
The value of $k_0$ is consistent, within $2\sigma$, with the expected value of 0.35, providing 
empirical support, from the basis of RM experiments, for the geometrically thin disk model.

The analysis shows high uncertainty on the value of $\mathdotM_c$. This is due to the limited 
numbers of objects with $\mathdotM>13.8$ (13 objects). The precision we 
can determine the values of $\mathdotM_c$ and $\Ysat$ would most likely improve when large
samples of high accretion rates become available.

\subsubsection{Slim Accretion Disks}
To understand the dependence of the radius-mass parameter on the accretion 
rates, we need to consider the properties of slim accretion disks.
Such disks are the result of the increasing mass accretion 
flow through a thin accretion disk which leads to a build-up of radiation pressure close 
to the BH, the thickening of the disk, and the trapping of the photons that are trying to 
escape from these regions. Such systems have been discussed in numerous papers (e.g., 
Abramowicz et al. 1988; Wang \& Zhou 1999; Mineshige et al. 2000; Watarai et al. 2001;
Sadowski 2009, and references therein). The general result obtained from these studies  
is a simple dependence of the bolometric luminosity of the system linearly on the BH mass 
and only a weak logarithmic dependence on the mass accretion rate (so called ``saturated 
luminosity"). The characteristic saturated bolometric luminosity can be expressed as
(Mineshige et al. 2000; see also Watarai et al. 2001)
\begin{equation}
L_{\rm Bol}=\ell_0\dotmfactor L_{\rm Edd}\,\, ~~~
        ({\rm for}~\mathdotM\ge \mathdotM_c),
\end{equation}
where $\ell_0=2\sim3$ and $\mathdotM_c\sim 20$. The uncertainties on $\ell_0$ and $\mathdotM_c$ 
depend on the details of the vertically-averaged equations of slim disks.
Calculations of emergent spectra of slim disks show that $L_{5100}\propto \mathdotM^{2/3}\bhm^{4/3}$
very similar to the SS73 disk, whereas the hydrogen ionizing luminosity 
($L_{\rm ion}=\int_{13.6{\rm eV}}^{\infty} L_{\epsilon}d\epsilon$, where $\epsilon$ is 
photon energy) increases with $\mathdotM$ much slower than $L_{5100}$ (Wang et al. 1999; 
Mineshige et al. 2000; Watarai et al. 2001; Wang et al. 2014b) and, for small to intermediate size 
BHs, $L_{\rm ion}\approx L_{\rm Bol}$ for large $\mathdotM$, showing a characteristic saturation 
of $L_{\rm ion}$ with $\mathdotM$. A canonic spectral energy distribution,  
$L_{\epsilon}\propto \epsilon^{-1}\exp(-\epsilon/\epsilon_0)$, is expected from the photon 
trapping part of slim disks, where $\epsilon_0$ is 
the cutoff energy determined by the maximum temperature of the disk. Such a
SED follows from a temperature distribution of $T_{\rm eff}\propto R^{-1/2}$, where $R$ is the
radius. This is significantly flatter than that in the SS73 disk 
(Wang \& Zhou 1999; Mineshige et al. 2000). 

The calculations of the structure and emitted spectrum of slim accretion disks are very 
challenging because of the complicated coupling between structure and radiation transfer.
The classical model (Abramowicz et al. 1988) uses the vertically-averaged equations (their 
validity regimes were discussed by Gu \& Lu 2007 and Cao \& Gu 2015), which probably underestimate
photon trapping effects (Ohsuga et al. 2002). The model neglects the time delay between 
energy generation around mid-plane and photon escaping from the surface of the disks. 
Therefore, both the saturated luminosity and the transition accretion rates are not 
well determined by theory. Numerical simulations of slim disks show different, perhaps 
more realistic properties, but the agreement between 2D fully general relativistic 
(Sadowski et al. 2014) and 3D magnetohydrodynamic calculations (Jiang et al. 2014) is 
not very good. In particular, in 3D simulations, the inclusion of vertical transportation 
of radiation fluxes driven by magnetic buoyancy significantly increases the output power 
of radiation from slim disks. However, such a process could be greatly suppressed by the 
fast radial motion of the accretion flows so that the radiation remains at a level of 
low efficiency as the classical model (Abramowicz et al. 2013). The detailed 
discussion of these theoretical issues is beyond the scope of this paper. Here we focus on
the attempt to use our new sample in order to qualitatively compare with the saturated 
luminosity and the transition accretion rates. Quantitative comparison with models 
of slim disks is given in a forthcoming paper.

The best measured value of $\Ysat$ (Equation 12) results in two expressions for the
critical BLR size,
\begin{equation}
\rrhb=3.3^{+0.6}_{-0.5}\times 10^4~m_7^{0.29};~~{\rm or}~~
R_{\rm crit}({\rm H\beta})=18.8^{+3.4}_{-2.8} ~m_7^{1.29}~{\rm ltd},
\end{equation}
which are independent of the accretion rates for AGNs with $\mathdotM\gtrsim \mathdotM_c$. 
In the context of photoionization, the critical BLR size corresponds to a saturation 
of the ionizing luminosity, and/or the bolometric luminosity. In particular, 
the small scatter of $\Ysat$ indicates that the scatter 
of the saturated luminosity is small in the current SEAMBH sample. Obviously, the 
$\Ythin-$saturation qualitatively supports the saturated luminosity shown by Equation (13), 
but Equation (14) provides a quantitative way to test theoretical models and numerical 
simulations of slim disks. The ultimate test for this finding will be performed once there 
are enough objects with the same BH mass with directly measured $\rhb$.

While the uncertainty on the preferred value of the transition accretion rate is 
large, because of the small number of objects, it is clear that
the value obtained from fitting our sample is considerably lower than $\mathdotM_c=50$ 
preferred by Mineshige et al. (2000), but covers $\mathdotM_c=20$ within the error bar 
given by Watarai et al. (2001).  
It has been suggested that the classical slim disk models may have 
underestimated the effects of photon trapping.
In fact, $\mathdotM_c \lesssim 10$ is predicted by model A of Ohsuga et al. (2002; their 
Figure 1). In this model, it is assumed that most of dissipated gravitational energy 
is released in the mid-plane and the vertical structure of gas density is presumed;
however, this assumption is not self-consistent. 
We also note that the minimum accretion rates of $\mathdotM_{\rm min}=3$ used by us 
to define SEAMBHs is outside the range found here. 
However, there must be a smooth transition from thin to slim disks and a relatively
large range in $\mathdotM$ from linear dependence on $\mathdotM^{0.35}$ to full saturation. 

In summary, the $\Ythin-$saturation measured from RM experiments provides a new tool 
to test the physics of accretion onto black holes. We find that the
current data support both sides of disk paradigm, from thin disks with the expected
dependence on mass and accretion rate, to slim disks with their expected saturation.

\subsubsection{The Maximum Size of the BLR}
We can also calculate the maximum expected dimensionless radius, $\rrhb$, using 
pure observational considerations.
For this we use Equation (3) and the definition of  $\rrhb$ from Equation (9). This gives,
$r_{_{\rm H\beta}}=\rblr/\Rg=\fblr^{-1}\left(c/V_{\rm FWHM}\right)^2$. This allows us to 
use the {\it smallest observed}  H$\beta$ line width to derive a mass-independent maximum
dimensionless size for all RM sources,
\begin{equation}
r_{_{\rm H\beta}}^{\rm max}=\fblr^{-1}\left(\frac{c}{V_{\rm min}}\right)^2
                           =9\times 10^4~\fblr^{-1} V_{\rm min,3}^{-2}, 
\end{equation}
where $V_{\rm min,3}=V_{\rm min}/10^3\kms$ is the minimum observed FWHM(H$\beta$).

Our sample includes one source, Mrk\,493, with FWHM(H$\beta) \sim 780\kms$. This width is, 
arguably, the smallest FWHM observed so far (see e.g.,  Zhou et al. 2006; Hu et al. 2008b). 
This translates to $r_{_{\rm H\beta}}^{\rm max} \simeq 1.5 \times 10^5$. As explained, 
this maximum gravitationally scaled size is independent of the BH mass. For systems powered 
by thin accretion disks we use the definition of $r_{\rm crit}=2.7\times 10^4~m_7^{-0.29}$ 
(by inserting $\mathdotM=3$ into Equation 9) to find the BH mass requires 
to get such a scaled radius, $m_7 \simeq 0.003$. The data in Table 6 shows that for 
Mrk\,493 $m_7 \simeq 0.14$ and $\mathdotM \simeq 76$, which indicates that this AGN is not 
powered by a thin disk. It is therefore not surprising that the two mass estimates are not 
in agreement.

\subsection{Slim Disks and the Size of the Torus}
The emissivity-weighted radius of the BLR measured by RM experiments is directly related to 
dust in the outskirts of the BLR, probably in a toroidal-shaped  structure (the ``central 
torus''). As argued in Barvainis (1987) and Netzer \& Laor (1993), the innermost boundary 
of the torus is set by the dust sublimation temperature which introduces a simple 
$R_{\rm dust} \propto L_{\rm Bol}^{1/2}$ relationship. Recent dust RM experiments (Suganuma 
et al. 2006; Koshida et al. 2014) provided $R_{\rm dust}$ in 17 sources by following the 
response of the dust emission in the $K$-band to the variable $V$-band continuum. The results 
are in very close agreement with the estimated graphite-grain sublimation radius assuming a 
mean grain size of about 0.05$\mu$m. Mor \& Netzer (2012) estimated this radius to be, 
$R_{\rm dust}=0.5 L_{46}^{1/2}T_{1800}^{2.8}$pc, where $L_{46}=L_{\rm Bol}/10^{46}\ergs$ and  
$T_{1800}=T_{\rm sub}/1800$K is the sublimation temperature. 

A comparison of the Koshida et al. (2014) observations and the ones presented here for 
$\mathdotM<3$ sources shows that, on average, $R_{\rm dust}/\rblr \simeq 4$. For $\mathdotM\ge 3$
objects, we find $R_{\rm dust}/\rblr \simeq 7$ (based only on four objects). Since the numbers 
are too small to derive a meaningful correlations for the two accretion rate sub-groups, we focus 
on the comparison of the measured BLR and torus sizes.

Figure 7  compares the dust and H$\beta$ radii as a function of $\mathdotM$ in two different 
ways. The left panel (7a) shows the dimensionless radii of the BLR and 
the torus inner walls for 14 objects in the Koshida et al. (2014) sample for which we have 
measured H$\beta$ lags. The right panel (7b) shows the ratio of  the two radii for all these 
objects. The surprising result is the unexpectedly large scatter in this ratio. While H$\beta$ 
and the hot dust follow the continuum luminosity in basically an identical way (size proportional 
to $L^{1/2}$), and differ only in their normalization, this does not result in similar ratios 
between the two distances for all sources. In fact, the ratios we measure range from about 1.4 
for Mrk\,590 to about 10 for Mrk\,335 and NGC\,4593. The median ratio, shown in the diagram, 
is about 4. The large scatter and the very small number of sources with measured dust lags 
prevent us from obtaining a meaningful relationship between $R_{\rm dust}$ and  $\mathdotM$.

There are several interesting ideas about the location of the innermost walls of the torus around 
slim accretion disks (e.g., Kawakatu \& Ohsuga 2011; Kawaguchi 2013). In particular, the large 
attenuation of the slim disk radiation at high polar angles (Wang et al. 2014b) would predict 
a small $R_{\rm dust}$ at these locations, perhaps as small as the self-gravity radius of the 
disk. This is not evident in the present data which suggests that the geometry of the inner torus 
walls is more complicated than the simplistic model suggested in various papers. 

\subsection{SEAMBHs and Cosmology}
The new results presented here suggest that the $\rblrl$ relationship for AGNs requires an 
additional parameter. In particular, the use of the earlier correlations that are based on 
low accretion rate sources, tend to over-estimate the BLR size of high$-\mathdotM$ sources 
and hence their BH mass. The shorter H$\beta$ lags found here indicate that these objects 
have smaller BLR size than the ones obtained by applying the old $\rblrl$ relationships. This 
implies smaller BH mass and, therefore, higher Eddington ratios. Recent work by  Kelly \& Shen 
(2013), Nobuta et al. (2012), and Netzer \& Trakhtenbrot (2014) all show the large fraction of 
high $\er$ sources at high redshifts. This number is likely to increase given the necessary 
correction to the BH mass found here. 

There are several ways of using AGNs for cosmology. These include
using the measured diameters of various BLRs (Elvis \& Karovska 2002), the empirical
$\rblrl$ relation (Horne et al. 2003; Watson et al. 2011; Czerny et al. 2013),
torus inner edge reverberation (Yoshii et al. 2014; H\"onig et al. 2014), and RMS
X-ray spectra (La Franca et al. 2014). The use of SEAMBHs as cosmological distance 
indicators has been suggested by Wang et al. (2013) and explained in great detail in 
Paper II where we examined the consequences of increasing $\mathdotM$ that leads to an 
almost constant $L_{\rm Bol}/\bhm$. Marziani \& Sulentic (2014) used a similar approach 
based on a sample of highly accreting AGNs identified by their 4D eigenvector 
1 (see also King et al. 2014). While we used RM-based BH mass measurements (Paper II), 
they used the standard single-epoch method, combined with photoionization modeling, 
to estimate $\rhb$

The usefulness of the present method for cosmology depends on the number of SEAMBHs with 
accurately measured $\bhm$ at high redshifts. As shown above, the reduction in the BLR 
size in SEAMBHs compared to the size in low accretion rate sources (used so far to estimate 
BH mass) depends on $\mathdotM$ and can reach a factor of $3\sim 4$ for  $\mathdotM=100$
(Figure 4a and Equation 7). It is reasonable to expect that higher values of $\mathdotM$ 
are associated with an even greater reduction relative to expressions currently used 
in the literature. At redshifts $z=1\sim 2$, the reduction is more significant than the 
increase in the measured lag due to the cosmological dilution factor ($1+z$)  
suggesting that many SEAMBHs at 
high redshifts, even those with higher luminosity, can be monitored, successfully, and 
their $\rblr$ measured, during one observing season, given adequate observing facilities.

\section{Conclusions}
We have completed two years of spectroscopic monitoring of 18 AGNs that are high-accretion 
candidates. The second year observations resulted in significant H$\beta$ lag measurements 
of five AGNs, all of which are identified as SEAMBHs. The highest accretion rate reaches 
$\mathdotM\gtrsim 200$ in a few objects. The main results of our study can be summarized as 
follows: 
{\begin{itemize}
\item
For a given $L_{5100}$, the SEAMBHs monitored by us generally have shorter H$\beta$ lags 
by a factor of up to $\sim 3-4$ compared with lower accretion rate AGNs. The reduction gets 
larger with increasing dimensionless accretion rates.

\item
We considered several possible explanations for the new results and suggest 
that they may be related to the strong anisotropy and self-shadowing of the central slim disk 
radiation. We defined a new radius-mass parameter, $\Ythin=m_7^{0.29}\rrhb $ and demonstrated 
that it reaches saturation when the accretion rate exceeds a critical value of 
$\mathdotM_c=13.8_{-8.1}^{+19.6}$. This transition accretion rate is significantly smaller 
than  the one predicted by classical slim disks model, but is in better agreement with modified 
slim disk models.

\item
The shorter H$\beta$ lags imply that the number of SEAMBHs in the universe probably has been
underestimated in earlier works that did not take into account the smaller BLR in such sources.
It also helps to obtain lag measurements in a shorter time, thus making the application of 
SEAMBHs to cosmology more feasible.  
\end{itemize}

Our observed sample, combined with SEAMBHs obtained from the literature, is too small 
to put stronger constraints on the radius-mass saturation parameter. Larger SEAMBH samples 
with a large dynamic range in $\mathdotM$, will provide better understanding of the nature 
of transition from thin to slim disk systems. Better and more accurate BH mass measurements, 
based on improved numerical tools such as the MCMC technique as applied to velocity-resolved 
RM observations (Pancoast et al. 2011; Li et al. 2013), will help too.
Finally, we are aware of the limitations of the current theory and numerical simulations
of slim accretion disks. Present super-Eddington accretion models are still
missing important physical ingredients and the observations presented here will be 
compared with more detailed calculations once these become available.

\acknowledgements{We are grateful to an anonymous referee for
many useful and thoughtful comments and suggestions that helped us to improve the paper.
We acknowledge the support of the staff of the Lijiang 2.4m telescope. 
Funding for the telescope has been provided by CAS and the People's Government of Yunnan 
Province. This research is supported by the Strategic Priority Research Program - The 
Emergence of Cosmological Structures of the Chinese Academy of Sciences, Grant No. XDB09000000, 
by NSFC grants NSFC-11173023, -11133006, -11373024, 
-11233003 and -11473002, and by Israel-China ISF-NSFC grant 83/13. }

\clearpage

\begin{deluxetable}{lllllcccccrrc}
\rotate
\tablecolumns{13}
\tablewidth{0pt}
\setlength{\tabcolsep}{4pt}
\tablecaption{The SEAMBH project: targets and observations}
\tabletypesize{\scriptsize}
\tablehead{
\colhead{Object}                &
\colhead{$\alpha_{2000}$}       &
\colhead{$\delta_{2000}$}       &
\colhead{redshift}              &
\colhead{monitoring period}     &
\colhead{$N_{\rm spec}$}        &
\multicolumn{3}{c}{S/N}         &
\colhead{}                      &
\multicolumn{2}{c}{Comparison stars} &
\colhead{Note on $\tauhb$ }                  \\ \cline{11-12} \cline{7-9}
\colhead{}                      &
\colhead{}                      &
\colhead{}                      &
\colhead{}                      &
\colhead{}                      &
\colhead{}                      &
\colhead{$m_{r^{\prime}}$(Lj)}               &
\colhead{$m_{r^{\prime}}$(WO)}      &
\colhead{Spec}                  &
\colhead{}                      &
\colhead{$R_*$}                 &
\colhead{P.A.}                  &
\colhead{}
}
\startdata
\multicolumn{13}{c}{First phase: SEAMBH2012 sample}\\ \hline
Mrk 335          & 00 06 19.5 & $+$20 12 10 & 0.0258 & Oct., 2012 $-$ Feb., 2013& 91 & 50 &$\cdots$& 21 & & $80^{\pp}.7$  & $174.5^{\circ}$  &Yes  \\
Mrk 1044         & 02 30 05.5 & $-$08 59 53 & 0.0165 & Oct., 2012 $-$ Feb., 2013& 77 & 91 &$\cdots$& 21 & & $207^{\pp}.0$ & $-143.0^{\circ}$ &Yes \\
IRAS 04416+1215  & 04 44 28.8 & $+$12 21 12 & 0.0889 & Oct., 2012 $-$ Mar., 2013& 92 & 12 &$\cdots$& 31 & & $137^{\pp}.9$ & $-55.0^{\circ}$  & Yes \\
Mrk 382          & 07 55 25.3 & $+$39 11 10 & 0.0337 & Oct., 2012 $-$ May., 2013&123 & 25 &$\cdots$&  3 & & $198^{\pp}.4$ & $-24.6^{\circ}$  &Yes  \\
Mrk 142          & 10 25 31.3 & $+$51 40 35 & 0.0449 & Nov., 2012 $-$ Apr., 2013&119 & 45 &$\cdots$& 10 & & $113^{\pp}.1$ & $155.2^{\circ}$  &Yes  \\
MCG $+06-26-012$ & 11 39 13.9 & $+$33 55 51 & 0.0328 & Jan., 2013 $-$ Jun., 2013& 34 & 24 &$\cdots$& 37 & & $204^{\pp}.3$ & $46.1^{\circ}$   &Yes \\
Mrk 42           & 11 53 41.8 & $+$46 12 43 & 0.0246 & Jan., 2013 $-$ Apr., 2013& 53 & 24 &$\cdots$& 15 & & $234^{\pp}.4$ & $33.8^{\circ}$   &No  \\
IRAS F12397+3333 & 12 42 10.6 & $+$33 17 03 & 0.0435 & Jan., 2013 $-$ May., 2013& 51 & 53 &$\cdots$& 32 & & $189^{\pp}.0$ & $130.0^{\circ}$  &Yes \\
Mrk 486          & 15 36 38.3 & $+$54 33 33 & 0.0389 & Mar., 2013 $-$ Jul., 2013& 45 &100 &$\cdots$& 44 & & $193^{\pp}.8$ & $-167.0^{\circ}$ &Yes \\
Mrk 493          & 15 59 09.6 & $+$35 01 47 & 0.0313 & Apr., 2013 $-$ Jun., 2013& 27 & 40 &$\cdots$& 45 & & $155^{\pp}.3$ & $98.5^{\circ}$   & Yes \\ \hline
\multicolumn{13}{c}{Second phase: SEAMBH2013 sample}\\ \hline
SDSS J075101.42+291419.1 & 07 51 01.4 & $+$29 14 19 & 0.1208 & Nov., 2013 $-$ May., 2014 & 38 &50&50&40& & $133^{\pp}.3$ & $-41.3^{\circ}$  & Yes \\
SDSS J080101.41+184840.7 & 08 01 01.4 & $+$18 48 40 & 0.1396 & Nov., 2013 $-$ Apr., 2014 & 34 &50&50&50& & $118^{\pp}.8$ & $-98.2^{\circ}$  & Yes \\
SDSS J080131.58+354436.4 & 08 01 31.6 & $+$35 44 36 & 0.1786 & Nov., 2013 $-$ Apr., 2014 & 31 &50&50&30& & $100^{\pp}.0$ & $145.2^{\circ}$  & uncertain \\
SDSS J081441.91+212918.5 & 08 14 41.9 & $+$21 29 19 & 0.1628 & Nov., 2013 $-$ May., 2014 & 34 &30&30&30& & $79^{\pp}.0$ & $73.9^{\circ}$    & Yes \\
SDSS J081456.10+532533.5 & 08 14 56.1 & $+$53 25 34 & 0.1197 & Nov., 2013 $-$ Apr., 2014 & 27 &40&...&50& & $164^{\pp}.5$ & $-172.9^{\circ}$ & Yes \\
SDSS J093922.89+370943.9 & 09 39 22.9 & $+$37 09 44 & 0.1859 & Nov., 2013 $-$ Jun., 2014 & 26 &40&30&30& & $175^{\pp}.1$ & $-139.0^{\circ}$ & Yes \\
SDSS J094422.13+103739.7 & 09 44 22.1 & $+$10 37 40 & 0.2410 & Jan., 2014 $-$ Apr., 2014 & 17 &30&30&15& & $109^{\pp}.3$ & $170.1^{\circ}$  & No \\
SDSS J100055.71+314001.2 & 10 00 55.7 & $+$31 40 01 & 0.1948 & Nov., 2013 $-$ Jun., 2014 & 26&100&15&40& & $151^{\pp}.2$ & $116.8^{\circ}$  & No \\
\enddata
\tablecomments{\footnotesize
We denote the samples monitored during the 2012--2013 and 
  2013--2014 observing seasons as SEAMBH2012 and SEAMBH2013, respectively. 
The SEAMBH2012 group is described in Papers I, II and III. 
No observational points were removed from the SEAMBH2012 sample due to S/N ratios.
$N_{\rm spec}$ is the numbers of spectroscopic epochs, $R_*$ is the angular distance
between the object and the comparison star and PA the position angle from the AGN to the
comparison star. The last column contains notes on the H$\beta$ time lags: ``Yes'' means 
significant lag, ``No" lags that could not be measured  
and ``uncertain" refers to the case of J080131 described in the text. $m_{r^{\prime}}$(Lj/WO) are 
referred to photometry at Lijiang Station of Yunnan Observatory and Wise Observatory. 
We calculated S/N ratios of the SEAMBH2012 and SEAMBH2013 samples from light curves. We
removed those points of SEAMBH2013 sample 
in poor observing conditions in light of the lowest S/N ratios listed here.
Only a few points with extremely large error bars were removed in SEAMBH2012 sample (see details in 
Papers I and II).  Objects marked with 
``$\cdots$'' were not observed at the Wise observatory.
}
\end{deluxetable}

\clearpage

\begin{deluxetable}{rlcrlcrlcrlcrlcrl}
\rotate
  \tablecolumns{17}
  \setlength{\tabcolsep}{3pt}
  \tablewidth{0pc}
  \tablecaption{Light curves of J075101 and J080101}
  \tabletypesize{\scriptsize}
  \tablehead{
      \multicolumn{8}{c}{J075101}        &
      \colhead{}                         &
      \multicolumn{8}{c}{J080101}        \\ \cline{1-8}\cline{10-17}
      \multicolumn{2}{c}{Continuum}      &
      \colhead{}                         &
      \multicolumn{2}{c}{Combined Continuum}     &
      \colhead{}                         &
      \multicolumn{2}{c}{Line}               &
      \colhead{}                         &
      \multicolumn{2}{c}{Continuum}      &
      \colhead{}                         &
      \multicolumn{2}{c}{Combined Continuum}     &
      \colhead{}                         &
      \multicolumn{2}{c}{Line}               \\ \cline{1-2}\cline{4-5}\cline{7-8}\cline{10-11}\cline{13-14}\cline{16-17}
      \colhead{JD}                       &
      \colhead{$F_{5100}$}               &
      \colhead{}                              &
      \colhead{JD}                       &
      \colhead{$F_{5100}$}               &
      \colhead{}                              &
      \colhead{JD}                       &
      \colhead{$F_{\rm H\beta}$}         &
      \colhead{}                              &
      \colhead{JD}                       &
      \colhead{$F_{5100}$}               &
      \colhead{}                              &
      \colhead{JD}                       &
      \colhead{$F_{5100}$}               &
      \colhead{}                              &
      \colhead{JD}                       &
      \colhead{$F_{\rm H\beta}$}
         }
\startdata
  3.466 & $ 6.219\pm 0.169$W & &    3.466 & $ 6.219\pm 0.169$ & &   10.277 & $ 4.312\pm 0.055$ & &   13.310 & $ 6.835\pm 0.089$L & &   13.310 & $ 6.835\pm 0.089$ & &   16.257 & $ 6.696\pm 0.076$ \\
  8.588 & $ 5.850\pm 0.181$W & &    8.588 & $ 5.850\pm 0.181$ & &   13.247 & $ 4.413\pm 0.074$ & &   16.257 & $ 7.120\pm 0.051$S & &   16.257 & $ 7.120\pm 0.051$ & &   21.430 & $ 6.484\pm 0.055$ \\
 10.277 & $ 6.325\pm 0.046$S & &   10.277 & $ 6.325\pm 0.046$ & &   15.248 & $ 4.508\pm 0.091$ & &   21.430 & $ 6.834\pm 0.046$S & &   21.440 & $ 6.793\pm 0.057$ & &   24.365 & $ 6.294\pm 0.058$ \\
 13.247 & $ 6.559\pm 0.075$S & &   13.256 & $ 6.455\pm 0.146$ & &   22.222 & $ 4.019\pm 0.053$ & &   21.449 & $ 6.753\pm 0.085$L & &   24.377 & $ 6.803\pm 0.064$ & &   29.270 & $ 6.330\pm 0.051$ \\
 13.265 & $ 6.352\pm 0.103$L & &   15.248 & $ 7.155\pm 0.074$ & &   25.255 & $ 3.699\pm 0.058$ & &   24.365 & $ 6.758\pm 0.076$S & &   29.283 & $ 6.866\pm 0.060$ & &   33.432 & $ 6.485\pm 0.062$ \\
  \enddata
  \tablecomments{\footnotesize
      JD: Julian dates from 2,456,600; $F_{5100}$ and $F_{\rm H\beta}$ are fluxes at $(1+z)5100$\AA\, 
      and H$\beta$ emission lines in units of $10^{-16}{\rm erg~s^{-1}~cm^{-2}~\AA^{-1}}$ and 
      $10^{-14}{\rm erg~s^{-1}~cm^{-2}}$.   L: Photometry from Lijiang. S: Spectrum from Lijiang. W: Photometry from Wise.      
      (This table is available in its entirety in a machine-readable form in the online journal. A portion is shown here for guidance regarding its form and content.)
      }
\end{deluxetable}

\clearpage

\begin{deluxetable}{rlcrlcrlcrlcrlcrl}
\rotate
  \tablecolumns{17}
  \setlength{\tabcolsep}{3pt}
  \tablewidth{0pc}
  \tablecaption{Light curves of J080131 and J081441}
  \tabletypesize{\scriptsize}
  \tablehead{
      \multicolumn{8}{c}{J080131}        &
      \colhead{}                         &
      \multicolumn{8}{c}{J081441}        \\ \cline{1-8}\cline{10-17}
      \multicolumn{2}{c}{Continuum}      &
      \colhead{}                         &
      \multicolumn{2}{c}{Combined Continuum}     &
      \colhead{}                         &
      \multicolumn{2}{c}{Line}               &
      \colhead{}                         &
      \multicolumn{2}{c}{Continuum}      &
      \colhead{}                         &
      \multicolumn{2}{c}{Combined Continuum}     &
      \colhead{}                         &
      \multicolumn{2}{c}{Line}               \\ \cline{1-2}\cline{4-5}\cline{7-8}\cline{10-11}\cline{13-14}\cline{16-17}
      \colhead{JD}                       &
      \colhead{$F_{5100}$}               &
      \colhead{}                              &
      \colhead{JD}                       &
      \colhead{$F_{5100}$}               &
      \colhead{}                              &
      \colhead{JD}                       &
      \colhead{$F_{\rm H\beta}$}         &
      \colhead{}                              &
      \colhead{JD}                       &
      \colhead{$F_{5100}$}               &
      \colhead{}                              &
      \colhead{JD}                       &
      \colhead{$F_{5100}$}               &
      \colhead{}                              &
      \colhead{JD}                       &
      \colhead{$F_{\rm H\beta}$}
         }
\startdata
  3.478 & $ 2.146\pm 0.072$W & &    3.478 & $ 2.146\pm 0.072$ & &    6.296 & $ 1.301\pm 0.022$ & &    3.490 & $ 2.720\pm 0.042$W & &    3.490 & $ 2.720\pm 0.042$ & &   18.318 & $ 2.590\pm 0.054$ \\
  6.296 & $ 2.230\pm 0.020$S & &    6.311 & $ 2.326\pm 0.136$ & &    7.336 & $ 1.318\pm 0.021$ & &    8.612 & $ 2.725\pm 0.036$W & &    8.612 & $ 2.725\pm 0.036$ & &   23.290 & $ 2.856\pm 0.035$ \\
  6.325 & $ 2.422\pm 0.029$L & &    7.351 & $ 2.258\pm 0.040$ & &    9.343 & $ 1.218\pm 0.020$ & &   18.318 & $ 2.502\pm 0.064$S & &   18.318 & $ 2.502\pm 0.064$ & &   30.409 & $ 2.782\pm 0.022$ \\
  7.336 & $ 2.287\pm 0.015$S & &    8.600 & $ 2.146\pm 0.068$ & &   12.401 & $ 1.221\pm 0.027$ & &   23.290 & $ 2.599\pm 0.028$S & &   23.309 & $ 2.626\pm 0.039$ & &   34.369 & $ 2.743\pm 0.028$ \\
  7.367 & $ 2.230\pm 0.029$L & &    9.343 & $ 2.241\pm 0.012$ & &   21.278 & $ 1.175\pm 0.024$ & &   23.328 & $ 2.654\pm 0.097$L & &   30.409 & $ 2.772\pm 0.010$ & &   37.301 & $ 2.728\pm 0.022$ \\
  \enddata
  \tablecomments{\footnotesize      
      This table is available in its entirety in a machine-readable form in the online journal. A portion is shown here for guidance regarding its form and content.
      }
\end{deluxetable}

\clearpage

\begin{deluxetable}{rlcrlcrlcrlcrlcrl}
\rotate
  \tablecolumns{17}
  \setlength{\tabcolsep}{3pt}
  \tablewidth{0pc}
  \tablecaption{Light curves of J081456 and J093922}
  \tabletypesize{\scriptsize}
  \tablehead{
      \multicolumn{8}{c}{J081456}        &
      \colhead{}                         &
      \multicolumn{8}{c}{J093922}        \\ \cline{1-8}\cline{10-17}
      \multicolumn{2}{c}{Continuum}      &
      \colhead{}                         &
      \multicolumn{2}{c}{Combined Continuum}     &
      \colhead{}                         &
      \multicolumn{2}{c}{Line}               &
      \colhead{}                         &
      \multicolumn{2}{c}{Continuum}      &
      \colhead{}                         &
      \multicolumn{2}{c}{Combined Continuum}     &
      \colhead{}                         &
      \multicolumn{2}{c}{Line} \\ \cline{1-2}\cline{4-5}\cline{7-8}\cline{10-11}\cline{13-14}\cline{16-17}
      \colhead{JD}                       &
      \colhead{$F_{5100}$}               &
      \colhead{}                              &
      \colhead{JD}                       &
      \colhead{$F_{5100}$}               &
      \colhead{}                              &
      \colhead{JD}                       &
      \colhead{$F_{\rm H\beta}$}         &
      \colhead{}                              &
      \colhead{JD}                       &
      \colhead{$F_{5100}$}               &
      \colhead{}                              &
      \colhead{JD}                       &
      \colhead{$F_{5100}$}               &
      \colhead{}                              &
      \colhead{JD}                       &
      \colhead{$F_{\rm H\beta}$}
         }
\startdata
  6.344 & $ 6.026\pm 0.028$S & &    6.349 & $ 6.089\pm 0.090$ & &    6.344 & $ 3.507\pm 0.041$ & &    3.589 & $ 2.673\pm 0.086$W & &    3.589 & $ 2.673\pm 0.086$ & &   17.372 & $ 1.300\pm 0.033$ \\
  6.354 & $ 6.152\pm 0.031$L & &    7.380 & $ 6.066\pm 0.030$ & &    7.380 & $ 3.441\pm 0.040$ & &   17.372 & $ 2.668\pm 0.014$S & &   17.372 & $ 2.668\pm 0.014$ & &   66.424 & $ 1.137\pm 0.029$ \\
  7.380 & $ 6.066\pm 0.030$S & &   10.430 & $ 6.083\pm 0.040$ & &   10.422 & $ 3.297\pm 0.039$ & &   50.596 & $ 2.554\pm 0.071$W & &   50.596 & $ 2.554\pm 0.071$ & &   68.304 & $ 1.070\pm 0.032$ \\
 10.422 & $ 6.054\pm 0.030$S & &   13.346 & $ 6.017\pm 0.046$ & &   19.319 & $ 3.184\pm 0.036$ & &   54.414 & $ 2.377\pm 0.078$W & &   54.414 & $ 2.377\pm 0.078$ & &   79.421 & $ 1.029\pm 0.038$ \\
 10.438 & $ 6.112\pm 0.024$L & &   19.319 & $ 5.974\pm 0.051$ & &   22.382 & $ 3.026\pm 0.046$ & &   56.441 & $ 2.537\pm 0.076$W & &   56.441 & $ 2.537\pm 0.076$ & &   81.239 & $ 1.140\pm 0.022$ \\
  \enddata
  \tablecomments{\footnotesize
      This table is available in its entirety in a machine-readable form in the online journal. A portion is shown here for guidance regarding its form and content.
      }
\end{deluxetable}

\begin{deluxetable}{rlcrlcrlcrlcrlcrl}
\rotate
  \tablecolumns{17}
  \setlength{\tabcolsep}{3pt}
  \tablewidth{0pc}
  \tablecaption{Light curves of J094422 and J100055}
  \tabletypesize{\scriptsize}
  \tablehead{
      \multicolumn{8}{c}{J094422}        &
      \colhead{}                         &
      \multicolumn{8}{c}{J100055}        \\ \cline{1-8}\cline{10-17}
      \multicolumn{2}{c}{Continuum}      &
      \colhead{}                         &
      \multicolumn{2}{c}{Combined Continuum}     &
      \colhead{}                         &
      \multicolumn{2}{c}{Line}               &
      \colhead{}                         &
      \multicolumn{2}{c}{Continuum}      &
      \colhead{}                         &
      \multicolumn{2}{c}{Combined Continuum}     &
      \colhead{}                         &
      \multicolumn{2}{c}{Line}               \\ \cline{1-2}\cline{4-5}\cline{7-8}\cline{10-11}\cline{13-14}\cline{16-17}
      \colhead{JD}                       &
      \colhead{$F_{5100}$}               &
      \colhead{}                              &
      \colhead{JD}                       &
      \colhead{$F_{5100}$}               &
      \colhead{}                              &
      \colhead{JD}                       &
      \colhead{$F_{\rm H\beta}$}         &
      \colhead{}                              &
      \colhead{JD}                       &
      \colhead{$F_{5100}$}               &
      \colhead{}                              &
      \colhead{JD}                       &
      \colhead{$F_{5100}$}               &
      \colhead{}                              &
      \colhead{JD}                       &
      \colhead{$F_{\rm H\beta}$}
         }
\startdata
  3.614 & $ 1.500\pm 0.038$W & &    3.614 & $ 1.500\pm 0.038$ & &   62.387 & $ 1.051\pm 0.020$ & &    9.512 & $ 2.921\pm 0.248$W & &    9.512 & $ 2.921\pm 0.248$ & &   18.388 & $ 0.697\pm 0.040$ \\
  9.476 & $ 1.405\pm 0.063$W & &    9.476 & $ 1.405\pm 0.063$ & &   67.447 & $ 1.043\pm 0.027$ & &   18.388 & $ 3.649\pm 0.042$S & &   18.388 & $ 3.649\pm 0.042$ & &   44.397 & $ 0.823\pm 0.046$ \\
 53.437 & $ 1.157\pm 0.035$W & &   53.437 & $ 1.157\pm 0.035$ & &   81.351 & $ 1.030\pm 0.029$ & &   43.551 & $ 2.932\pm 0.237$W & &   43.551 & $ 2.932\pm 0.237$ & &   64.438 & $ 0.747\pm 0.030$ \\
 56.469 & $ 1.216\pm 0.040$W & &   56.469 & $ 1.216\pm 0.040$ & &   84.394 & $ 0.964\pm 0.024$ & &   44.397 & $ 3.483\pm 0.038$S & &   44.397 & $ 3.483\pm 0.038$ & &   67.261 & $ 0.798\pm 0.029$ \\
 61.528 & $ 1.222\pm 0.040$W & &   61.528 & $ 1.222\pm 0.040$ & &   87.234 & $ 0.961\pm 0.018$ & &   45.611 & $ 3.203\pm 0.265$W & &   45.611 & $ 3.203\pm 0.265$ & &   71.420 & $ 0.802\pm 0.033$ \\
  \enddata
  \tablecomments{\footnotesize
      This table is available in its entirety in a machine-readable form in the online journal. A portion is shown here for guidance regarding its form and content.
      }
\end{deluxetable}

\clearpage
\renewcommand{\arraystretch}{1.5}
\begin{deluxetable}{lccccccrccc}
 \tablecolumns{9}
 \setlength{\tabcolsep}{3pt}
 \tablewidth{0pc}
 \tablecaption{H$\beta$ Reverberations of the SEAMBH Targets}
 \tabletypesize{\scriptsize}
 \tablehead{
     \colhead{Objects}                  &
     \colhead{$\tauhb$}                 &
     \colhead{FWHM}                     &
     \colhead{$\sigma_{\rm line}$}      &
     \colhead{$\log \left(\bhm/\sunm\right)$}&
     \colhead{$\log \mathdotM$}         &
     \colhead{$\log L_{5100}$}          &
     \colhead{$\log \Lhb$}              &
     \colhead{EW(H$\beta$)}             \\ \cline{2-9}
     \colhead{}                         &
     \colhead{(days)}                   &
     \colhead{($\kms$)}                 &
     \colhead{($\kms$)}                 &
     \colhead{}                         &
     \colhead{}                         &
     \colhead{($\ergs)$}                &
     \colhead{($\ergs)$}                &
     \colhead{(\AA)}
 }
\startdata
 \multicolumn{9}{c}{SEAMBH2012}\\ \hline
          Mrk 335  & $      8.7_{-  1.9}^{+  1.6} $ & $     2096\pm  170 $ & $1470\pm 50$ & $    6.87_{-0.14}^{+0.10} $ & $      1.28_{-  0.30}^{+ 0.37} $ & $    43.69\pm 0.06 $ & $    42.03\pm 0.06 $ & $    110.5\pm 22.3 $ \\ 
         Mrk 1044  & $     10.5_{-  2.7}^{+  3.3} $ & $     1178\pm   22 $ & $ 766\pm  8$ & $    6.45_{-0.13}^{+0.12} $ & $      1.22_{-  0.41}^{+ 0.40} $ & $    43.10\pm 0.10 $ & $    41.39\pm 0.09 $ & $    101.4\pm 31.9 $ \\ 
          Mrk 382  & $      7.5_{-  2.0}^{+  2.9} $ & $     1462\pm  296 $ & $ 840\pm 37$ & $    6.50_{-0.29}^{+0.19} $ & $      1.18_{-  0.53}^{+ 0.69} $ & $    43.12\pm 0.08 $ & $    41.01\pm 0.05 $ & $     39.6\pm  9.0 $ \\ 
          Mrk 142  & $      7.9_{-  1.1}^{+  1.2} $ & $     1588\pm   58 $ & $ 948\pm 12$ & $    6.59_{-0.07}^{+0.07} $ & $      1.65_{-  0.23}^{+ 0.23} $ & $    43.56\pm 0.06 $ & $    41.60\pm 0.04 $ & $     55.2\pm  9.5 $ \\ 
MCG +06-26-012$^a$  & $     24.0_{-  4.8}^{+  8.4} $ & $     1334\pm   80 $ & $ 785\pm 21$ & $    6.92_{-0.12}^{+0.14} $ & $     -0.34_{-  0.45}^{+ 0.37} $ & $    42.67\pm 0.11 $ & $    41.03\pm 0.06 $ & $    114.6\pm 32.5 $ \\ 
  IRAS F12397$^b$  & $      9.7_{-  1.8}^{+  5.5} $ & $     1802\pm  560 $ & $1150\pm122$ & $    6.79_{-0.45}^{+0.27} $ & $      2.26_{-  0.62}^{+ 0.98} $ & $    44.23\pm 0.05 $ & $    42.26\pm 0.04 $ & $     54.2\pm  8.4 $ \\ 
          Mrk 486  & $     23.7_{-  2.7}^{+  7.5} $ & $     1942\pm   67 $ & $1296\pm 23$ & $    7.24_{-0.06}^{+0.12} $ & $      0.55_{-  0.32}^{+ 0.20} $ & $    43.69\pm 0.05 $ & $    42.12\pm 0.04 $ & $    135.9\pm 20.3 $ \\ 
          Mrk 493  & $     11.6_{-  2.6}^{+  1.2} $ & $      778\pm   12 $ & $ 513\pm  5$ & $    6.14_{-0.11}^{+0.04} $ & $      1.88_{-  0.21}^{+ 0.33} $ & $    43.11\pm 0.08 $ & $    41.35\pm 0.05 $ & $     87.4\pm 18.1 $ \\ 
   IRAS 04416$^c$  & $     13.3_{-  1.4}^{+ 13.9} $ & $     1522\pm   44 $ & $1056\pm 29$ & $    6.78_{-0.06}^{+0.31} $ & $      2.63_{-  0.67}^{+ 0.16} $ & $    44.47\pm 0.03 $ & $    42.51\pm 0.02 $ & $     55.8\pm  4.7 $ \\  \hline
 \multicolumn{9}{c}{SEAMBH2013}\\ \hline
      SDSS J075101  & $     33.4_{-  5.6}^{+ 15.6} $ & $     1495\pm   67 $ & $1055\pm32$ & $    7.16_{-0.09}^{+0.17} $ & $      1.34_{-  0.41}^{+ 0.25} $ & $    44.12\pm 0.05 $ & $    42.25\pm 0.03 $ & $     68.1\pm  8.6 $ \\ 
      SDSS J080101  & $      8.3_{-  2.7}^{+  9.7} $ & $     1930\pm   18 $ & $1119\pm 3$ & $    6.78_{-0.17}^{+0.34} $ & $      2.33_{-  0.72}^{+ 0.39} $ & $    44.27\pm 0.03 $ & $    42.58\pm 0.02 $ & $    105.5\pm  8.3 $ \\ 
      SDSS J081441  & $     18.4_{-  8.4}^{+ 12.7} $ & $     1615\pm   22 $ & $1122\pm11$ & $    6.97_{-0.27}^{+0.23} $ & $      1.56_{-  0.57}^{+ 0.63} $ & $    44.01\pm 0.07 $ & $    42.42\pm 0.03 $ & $    132.0\pm 23.7 $ \\ 
      SDSS J081456  & $     24.3_{- 16.4}^{+  7.7} $ & $     2409\pm   61 $ & $1438\pm32$ & $    7.44_{-0.49}^{+0.12} $ & $      0.59_{-  0.30}^{+ 1.03} $ & $    43.99\pm 0.04 $ & $    42.15\pm 0.03 $ & $     74.4\pm  7.6 $ \\ 
      SDSS J093922  & $     11.9_{-  6.3}^{+  2.1} $ & $     1209\pm   16 $ & $ 835\pm11$ & $    6.53_{-0.33}^{+0.07} $ & $      2.54_{-  0.20}^{+ 0.71} $ & $    44.07\pm 0.04 $ & $    42.09\pm 0.04 $ & $     53.0\pm  6.7 $ \\ 
 \enddata
 \tablecomments{\footnotesize 
1: H$\beta$ lags of J080131 are not listed in this table, but given in the second paragraph of Section 3
in the main text. 2: All SEAMBH2012 measurements are taken from Paper III, but 5100\AA\, fluxes are from 
I and II. $^a$MCG +06$-$26$-$012 was selected as a super-Eddington candidate but later was identified 
to be a sub-Eddington accretor ($\mathdotM=0.46$). $^b$For IRAS F12397, we use the fluxes of the case 
with local absorption correction (see the details in Paper I and III). $^c$ The time lag of IRAS 04416 
cannot be obtained significantly using the integration method in Papers I and II, but
has been detected by the fitting procedures in Paper III.}
\end{deluxetable}

\clearpage
\renewcommand{\arraystretch}{1.5}
\begin{deluxetable}{lcccccccl}
\tablecolumns{9}
\setlength{\tabcolsep}{3pt}
\tablewidth{0pc}
\tablecaption{Reverberation mapping AGNs and results}
\tabletypesize{\scriptsize}
\tablehead{
    \colhead{Objects}                  &
    \colhead{$\tauhb$}                 &
    \colhead{FWHM}                     &
    \colhead{$\log\left(\bhm/\sunm\right)$} &
    \colhead{$\log\mathdotM$}&
    \colhead{$\log L_{5100}$}          &
    \colhead{$\log \Lhb$}              &
    \colhead{EW(H$\beta$)}             &
    \colhead{Ref.}                     \\ \cline{2-9}
    \colhead{}                         &
    \colhead{(days)}                   &
    \colhead{($\kms$)}                 &
    \colhead{}                         &
    \colhead{}                         &
    \colhead{($\ergs)$}                &
    \colhead{($\ergs)$}                &
    \colhead{(\AA)}                    &
    \colhead{}
}
\startdata
          Mrk 335  & $      8.7_{-  1.9}^{+  1.6} $ & $     2096\pm  170 $ & $    6.87_{-0.14}^{+0.10} $ & $      1.17_{-  0.30}^{+  0.31} $ & $    43.69\pm 0.06 $ & $    42.03\pm 0.06 $ & $    110.5\pm 22.3 $ & 1,2,3        \\ 
                   & $     16.8_{-  4.2}^{+  4.8} $ & $     1792\pm    3 $ & $    7.02_{-0.12}^{+0.11} $ & $      1.28_{-  0.29}^{+  0.30} $ & $    43.76\pm 0.06 $ & $    42.13\pm 0.06 $ & $ 119.7\pm 23.3 $ & 4,5,6,7      \\ 
                   & $     12.5_{-  5.5}^{+  6.6} $ & $     1679\pm    2 $ & $    6.84_{-0.25}^{+0.18} $ & $      1.39_{-  0.29}^{+  0.30} $ & $    43.84\pm 0.06 $ & $    42.18\pm 0.06 $ & $ 111.2\pm 21.1 $ & 4,5,6,7      \\ 
                   & $     14.3_{-  0.7}^{+  0.7} $ & $     1724\pm  236 $ & $    6.92_{-0.14}^{+0.11} $ & $      1.25_{-  0.29}^{+  0.30} $ & $    43.74\pm 0.06 $ & $    41.99\pm 0.07 $ & $     89.5\pm 19.5 $ & 4,8$^a$      \\ 
                   & $\bm{ 14.0_{-  3.4}^{+  4.6}}$ & $\bm{ 1707\pm   79}$ & $\bm{6.93_{-0.11}^{+0.10}}$ & $\bm{  1.27_{-  0.17}^{+  0.18}}$ & $\bm{43.76\pm 0.07}$ & $\bm{42.09\pm 0.09}$ & $\bm{108.2\pm 16.8}$ &              \\ 
      PG 0026+129  & $    111.0_{- 28.3}^{+ 24.1} $ & $     2544\pm   56 $ & $    8.15_{-0.13}^{+0.09} $ & $      0.65_{-  0.20}^{+  0.28} $ & $    44.97\pm 0.02 $ & $    42.93\pm 0.04 $ & $     46.2\pm  4.7 $ & 4,5,6,9      \\ 
      PG 0052+251  & $     89.8_{- 24.1}^{+ 24.5} $ & $     5008\pm   73 $ & $    8.64_{-0.14}^{+0.11} $ & $     -0.59_{-  0.25}^{+  0.31} $ & $    44.81\pm 0.03 $ & $    43.13\pm 0.05 $ & $    107.4\pm 14.8 $ & 4,5,6,9      \\ 
         Fairall9  & $     17.4_{-  4.3}^{+  3.2} $ & $     5999\pm   66 $ & $    8.09_{-0.12}^{+0.07} $ & $     -0.71_{-  0.21}^{+  0.31} $ & $    43.98\pm 0.04 $ & $    42.67\pm 0.04 $ & $    249.8\pm 32.0 $ & 4,5,6,10     \\ 
          Mrk 590  & $     20.7_{-  2.7}^{+  3.5} $ & $     2788\pm   29 $ & $    7.50_{-0.06}^{+0.07} $ & $     -0.22_{-  0.25}^{+  0.24} $ & $    43.59\pm 0.06 $ & $    41.92\pm 0.06 $ & $    107.0\pm 22.0 $ & 4,5,6,7      \\ 
                   & $     14.0_{-  8.8}^{+  8.5} $ & $     3729\pm  426 $ & $    7.58_{-0.48}^{+0.22} $ & $     -0.91_{-  0.30}^{+  0.28} $ & $    43.14\pm 0.09 $ & $    41.58\pm 0.16 $ & $    142.8\pm 62.2 $ & 4,5,6,7      \\ 
                   & $     29.2_{-  5.0}^{+  4.9} $ & $     2744\pm   79 $ & $    7.63_{-0.09}^{+0.07} $ & $     -0.54_{-  0.26}^{+  0.25} $ & $    43.38\pm 0.07 $ & $    41.75\pm 0.07 $ & $    119.7\pm 28.1 $ & 4,5,6,7      \\ 
                   & $     28.8_{-  4.2}^{+  3.6} $ & $     2500\pm   42 $ & $    7.55_{-0.07}^{+0.05} $ & $     -0.13_{-  0.25}^{+  0.24} $ & $    43.65\pm 0.06 $ & $    41.92\pm 0.07 $ & $     94.9\pm 20.7 $ & 4,5,6,7      \\ 
                   & $\bm{ 25.6_{-  5.3}^{+  6.5}}$ & $\bm{ 2716\pm  202}$ & $\bm{7.55_{-0.08}^{+0.07}}$ & $\bm{ -0.41_{-  0.36}^{+  0.36}}$ & $\bm{43.50\pm 0.21}$ & $\bm{41.85\pm 0.12}$ & $\bm{108.6\pm 20.2}$ &              \\ 
         Mrk 1044  & $     10.5_{-  2.7}^{+  3.3} $ & $     1178\pm   22 $ & $    6.45_{-0.13}^{+0.12} $ & $      1.22_{-  0.41}^{+  0.40} $ & $    43.10\pm 0.10 $ & $    41.39\pm 0.09 $ & $    101.4\pm 31.9 $ & 1,2,3        \\ 
           3C 120  & $     38.1_{- 15.3}^{+ 21.3} $ & $     2327\pm   48 $ & $    7.61_{-0.22}^{+0.19} $ & $      0.03_{-  0.37}^{+  0.37} $ & $    44.07\pm 0.05 $ & $    42.37\pm 0.06 $ & $    100.9\pm 18.3 $ & 4,5,6,7      \\ 
                   & $     25.9_{-  2.3}^{+  2.3} $ & $     3529\pm  176 $ & $    7.80_{-0.06}^{+0.05} $ & $     -0.17_{-  0.37}^{+  0.37} $ & $    43.94\pm 0.05 $ & $    42.36\pm 0.05 $ & $    135.9\pm 22.3 $ & 4,8$^a$      \\ 
                   & $\bm{ 26.2_{-  6.6}^{+  8.7}}$ & $\bm{ 2472\pm  729}$ & $\bm{7.79_{-0.15}^{+0.15}}$ & $\bm{ -0.07_{-  0.30}^{+  0.30}}$ & $\bm{44.00\pm 0.10}$ & $\bm{42.36\pm 0.04}$ & $\bm{118.8\pm 28.9}$ &              \\ 
       IRAS 04416  & $     13.3_{-  1.4}^{+ 13.9} $ & $     1522\pm   44 $ & $    6.78_{-0.06}^{+0.31} $ & $      2.63_{-  0.67}^{+  0.16} $ & $    44.47\pm 0.03 $ & $    42.51\pm 0.02 $ & $     55.8\pm  4.7 $ & 1,2,3        \\ 
          Ark 120  & $     47.1_{- 12.4}^{+  8.3} $ & $     6042\pm   35 $ & $    8.53_{-0.13}^{+0.07} $ & $     -1.48_{-  0.23}^{+  0.24} $ & $    43.98\pm 0.06 $ & $    42.60\pm 0.05 $ & $    211.5\pm 37.5 $ & 4,5,6,7      \\ 
                   & $     37.1_{-  5.4}^{+  4.8} $ & $     6246\pm   78 $ & $    8.45_{-0.07}^{+0.05} $ & $     -2.01_{-  0.27}^{+  0.27} $ & $    43.63\pm 0.08 $ & $    42.43\pm 0.07 $ & $    321.1\pm 77.6 $ & 4,5,6,7      \\ 
                   & $\bm{ 39.5_{-  7.8}^{+  8.5}}$ & $\bm{ 6077\pm  147}$ & $\bm{8.47_{-0.08}^{+0.07}}$ & $\bm{ -1.70_{-  0.41}^{+  0.41}}$ & $\bm{43.87\pm 0.25}$ & $\bm{42.54\pm 0.13}$ & $\bm{244.8\pm 80.3}$ &              \\ 
           Mrk 79  & $      9.0_{-  7.8}^{+  8.3} $ & $     5056\pm   85 $ & $    7.65_{-0.88}^{+0.28} $ & $     -0.75_{-  0.34}^{+  0.41} $ & $    43.63\pm 0.07 $ & $    41.89\pm 0.07 $ & $     92.4\pm 21.0 $ & 4,5,6,7      \\ 
                   & $     16.1_{-  6.6}^{+  6.6} $ & $     4760\pm   31 $ & $    7.85_{-0.23}^{+0.15} $ & $     -0.59_{-  0.34}^{+  0.41} $ & $    43.74\pm 0.07 $ & $    41.92\pm 0.08 $ & $     78.1\pm 18.3 $ & 4,5,6,7      \\ 
                   & $     16.0_{-  5.8}^{+  6.4} $ & $     4766\pm   71 $ & $    7.85_{-0.20}^{+0.15} $ & $     -0.70_{-  0.34}^{+  0.41} $ & $    43.66\pm 0.07 $ & $    41.89\pm 0.07 $ & $     86.0\pm 19.3 $ & 4,5,6,7      \\ 
                   & $\bm{ 15.6_{-  4.9}^{+  5.1}}$ & $\bm{ 4793\pm  145}$ & $\bm{7.84_{-0.16}^{+0.12}}$ & $\bm{ -0.68_{-  0.21}^{+  0.25}}$ & $\bm{43.68\pm 0.07}$ & $\bm{41.90\pm 0.05}$ & $\bm{ 85.4\pm 13.3}$ &              \\ 
     SDSS J075101  & $     33.4_{-  5.6}^{+ 15.6} $ & $     1495\pm   67 $ & $    7.16_{-0.09}^{+0.17} $ & $      1.34_{-  0.41}^{+  0.25} $ & $    44.12\pm 0.05 $ & $    42.25\pm 0.03 $ & $     68.1\pm  8.6 $ & 11           \\ 
          Mrk 382  & $      7.5_{-  2.0}^{+  2.9} $ & $     1462\pm  296 $ & $    6.50_{-0.29}^{+0.19} $ & $      1.18_{-  0.53}^{+  0.69} $ & $    43.12\pm 0.08 $ & $    41.01\pm 0.05 $ & $     39.6\pm  9.0 $ & 1,2,3        \\ 
     SDSS J080101  & $      8.3_{-  2.7}^{+  9.7} $ & $     1930\pm   18 $ & $    6.78_{-0.17}^{+0.34} $ & $      2.33_{-  0.72}^{+  0.39} $ & $    44.27\pm 0.03 $ & $    42.58\pm 0.02 $ & $    105.5\pm  8.3 $ & 11           \\ 
      PG 0804+761  & $    146.9_{- 18.9}^{+ 18.8} $ & $     3053\pm   38 $ & $    8.43_{-0.06}^{+0.05} $ & $      0.00_{-  0.13}^{+  0.15} $ & $    44.91\pm 0.02 $ & $    43.29\pm 0.03 $ & $    122.5\pm 10.3 $ & 4,5,6,9      \\ 
     SDSS J081441  & $     18.4_{-  8.4}^{+ 12.7} $ & $     1615\pm   22 $ & $    6.97_{-0.27}^{+0.23} $ & $      1.56_{-  0.57}^{+  0.63} $ & $    44.01\pm 0.07 $ & $    42.42\pm 0.03 $ & $    132.0\pm 23.7 $ & 11           \\ 
     SDSS J081456  & $     24.3_{- 16.4}^{+  7.7} $ & $     2409\pm   61 $ & $    7.44_{-0.49}^{+0.12} $ & $      0.59_{-  0.30}^{+  1.03} $ & $    43.99\pm 0.04 $ & $    42.15\pm 0.03 $ & $     74.4\pm  7.6 $ & 11           \\ 
      PG 0844+349  & $     32.3_{- 13.4}^{+ 13.7} $ & $     2694\pm   58 $ & $    7.66_{-0.23}^{+0.15} $ & $      0.50_{-  0.42}^{+  0.57} $ & $    44.22\pm 0.07 $ & $    42.56\pm 0.05 $ & $    111.2\pm 22.1 $ & 5,6,9,12     \\ 
          Mrk 110  & $     24.3_{-  8.3}^{+  5.5} $ & $     1543\pm    5 $ & $    7.05_{-0.18}^{+0.09} $ & $      0.81_{-  0.32}^{+  0.35} $ & $    43.68\pm 0.04 $ & $    42.12\pm 0.05 $ & $    139.6\pm 20.4 $ & 4,5,6,7      \\ 
                   & $     20.4_{-  6.3}^{+ 10.5} $ & $     1658\pm    3 $ & $    7.04_{-0.16}^{+0.18} $ & $      0.92_{-  0.32}^{+  0.34} $ & $    43.75\pm 0.04 $ & $    42.02\pm 0.05 $ & $     94.8\pm 14.7 $ & 4,5,6,7      \\ 
                   & $     33.3_{- 10.0}^{+ 14.9} $ & $     1600\pm   39 $ & $    7.22_{-0.16}^{+0.16} $ & $      0.58_{-  0.33}^{+  0.35} $ & $    43.53\pm 0.05 $ & $    41.97\pm 0.04 $ & $    139.4\pm 20.5 $ & 4,5,6,7      \\ 
                   & $\bm{ 25.6_{-  7.2}^{+  8.9}}$ & $\bm{ 1634\pm   83}$ & $\bm{7.10_{-0.14}^{+0.13}}$ & $\bm{  0.77_{-  0.25}^{+  0.26}}$ & $\bm{43.66\pm 0.12}$ & $\bm{42.03\pm 0.08}$ & $\bm{123.8\pm 29.1}$ &              \\ 
     SDSS J093922  & $     11.9_{-  6.3}^{+  2.1} $ & $     1209\pm   16 $ & $    6.53_{-0.33}^{+0.07} $ & $      2.54_{-  0.20}^{+  0.71} $ & $    44.07\pm 0.04 $ & $    42.09\pm 0.04 $ & $     53.0\pm  6.7 $ & 11           \\ 
      PG 0953+414  & $    150.1_{- 22.6}^{+ 21.6} $ & $     3071\pm   27 $ & $    8.44_{-0.07}^{+0.06} $ & $      0.39_{-  0.14}^{+  0.16} $ & $    45.19\pm 0.01 $ & $    43.29\pm 0.04 $ & $     64.7\pm  5.9 $ & 4,5,6,9      \\ 
         NGC 3227  & $      3.8_{-  0.8}^{+  0.8} $ & $     4112\pm  206 $ & $    7.09_{-0.12}^{+0.09} $ & $     -1.34_{-  0.36}^{+  0.38} $ & $    42.24\pm 0.11 $ & $    40.38\pm 0.10 $ & $     71.0\pm 23.6 $ & 4,13$^a$     \\ 
          Mrk 142  & $      7.9_{-  1.1}^{+  1.2} $ & $     1588\pm   58 $ & $    6.59_{-0.07}^{+0.07} $ & $      1.90_{-  0.86}^{+  0.85} $ & $    43.56\pm 0.06 $ & $    41.60\pm 0.04 $ & $     55.2\pm  9.5 $ & 1,2,3        \\ 
                   & $      2.7_{-  0.8}^{+  0.7} $ & $     1462\pm    2 $ & $    6.06_{-0.16}^{+0.10} $ & $      1.96_{-  0.82}^{+  0.82} $ & $    43.61\pm 0.04 $ & $    41.66\pm 0.05 $ & $ 57.6\pm  8.6 $ & 4,14         \\ 
                   & $\bm{  6.4_{-  3.4}^{+  7.3}}$ & $\bm{ 1462\pm   86}$ & $\bm{6.47_{-0.38}^{+0.38}}$ & $\bm{  1.93_{-  0.59}^{+  0.59}}$ & $\bm{43.59\pm 0.04}$ & $\bm{41.62\pm 0.06}$ & $\bm{ 56.6\pm  6.6}$ &              \\ 
         NGC 3516  & $     11.7_{-  1.5}^{+  1.0} $ & $     5384\pm  269 $ & $    7.82_{-0.08}^{+0.05} $ & $     -1.97_{-  0.52}^{+  0.41} $ & $    42.79\pm 0.20 $ & $    41.06\pm 0.18 $ & $     94.7\pm 59.2 $ & 4,13$^a$     \\ 
    SBS 1116+583A  & $      2.3_{-  0.5}^{+  0.6} $ & $     3668\pm  186 $ & $    6.78_{-0.12}^{+0.11} $ & $     -0.87_{-  0.71}^{+  0.51} $ & $    42.14\pm 0.23 $ & $    40.70\pm 0.07 $ & $    186.8\pm104.1 $ & 4,14         \\ 
          Arp 151  & $      4.0_{-  0.7}^{+  0.5} $ & $     3098\pm   69 $ & $    6.87_{-0.08}^{+0.05} $ & $     -0.44_{-  0.28}^{+  0.30} $ & $    42.55\pm 0.10 $ & $    40.95\pm 0.11 $ & $    130.0\pm 44.4 $ & 4,14         \\ 
         NGC 3783  & $     10.2_{-  2.3}^{+  3.3} $ & $     3770\pm   68 $ & $    7.45_{-0.11}^{+0.12} $ & $     -1.58_{-  0.59}^{+  0.45} $ & $    42.56\pm 0.18 $ & $    41.01\pm 0.18 $ & $    144.0\pm 83.7 $ & 4,5,6,15     \\ 
   MCG +06-26-012  & $     24.0_{-  4.8}^{+  8.4} $ & $     1334\pm   80 $ & $    6.92_{-0.12}^{+0.14} $ & $     -0.34_{-  0.45}^{+  0.37} $ & $    42.67\pm 0.11 $ & $    41.03\pm 0.06 $ & $    114.6\pm 32.5 $ & 1,2,3        \\ 
         Mrk 1310  & $      3.7_{-  0.6}^{+  0.6} $ & $     2409\pm   24 $ & $    6.62_{-0.08}^{+0.07} $ & $     -0.31_{-  0.39}^{+  0.35} $ & $    42.29\pm 0.14 $ & $    40.56\pm 0.10 $ & $     94.3\pm 38.2 $ & 4,14         \\ 
         NGC 4051  & $      1.9_{-  0.5}^{+  0.5} $ & $      851\pm  277 $ & $    5.42_{-0.53}^{+0.23} $ & $      1.59_{-  0.84}^{+  1.29} $ & $    41.96\pm 0.19 $ & $    40.19\pm 0.18 $ & $     86.8\pm 51.7 $ & 4,13$^a$     \\ 
         NGC 4151  & $      6.6_{-  0.8}^{+  1.1} $ & $     6371\pm  150 $ & $    7.72_{-0.06}^{+0.07} $ & $     -2.81_{-  0.57}^{+  0.37} $ & $    42.09\pm 0.21 $ & $    40.56\pm 0.20 $ & $    150.8\pm100.6 $ & 4,5,6,16     \\ 
      PG 1211+143  & $     93.8_{- 42.1}^{+ 25.6} $ & $     2012\pm   37 $ & $    7.87_{-0.26}^{+0.11} $ & $      0.84_{-  0.35}^{+  0.63} $ & $    44.73\pm 0.08 $ & $    43.02\pm 0.06 $ & $    100.2\pm 22.9 $ & 5,6,9,12     \\ 
          Mrk 202  & $      3.0_{-  1.1}^{+  1.7} $ & $     1471\pm   18 $ & $    6.11_{-0.20}^{+0.20} $ & $      0.66_{-  0.65}^{+  0.59} $ & $    42.26\pm 0.14 $ & $    40.40\pm 0.09 $ & $     70.6\pm 27.5 $ & 4,14         \\ 
         NGC 4253  & $      6.2_{-  1.2}^{+  1.6} $ & $     1609\pm   39 $ & $    6.49_{-0.10}^{+0.10} $ & $      0.36_{-  0.42}^{+  0.36} $ & $    42.57\pm 0.12 $ & $    40.77\pm 0.12 $ & $     81.1\pm 31.6 $ & 4,14         \\ 
      PG 1226+023  & $    306.8_{- 90.9}^{+ 68.5} $ & $     3509\pm   36 $ & $    8.87_{-0.15}^{+0.09} $ & $      0.70_{-  0.20}^{+  0.33} $ & $    45.96\pm 0.02 $ & $    44.13\pm 0.04 $ & $     74.6\pm  6.9 $ & 4,5,6,9      \\ 
      PG 1229+204  & $     37.8_{- 15.3}^{+ 27.6} $ & $     3828\pm   54 $ & $    8.03_{-0.23}^{+0.24} $ & $     -1.03_{-  0.55}^{+  0.52} $ & $    43.70\pm 0.05 $ & $    42.31\pm 0.06 $ & $    209.7\pm 38.3 $ & 4,5,6,9      \\ 
         NGC 4593  & $      3.7_{-  0.8}^{+  0.8} $ & $     5143\pm   16 $ & $    7.28_{-0.10}^{+0.08} $ & $     -0.73_{-  0.52}^{+  0.41} $ & $    42.87\pm 0.18 $ & $    41.17\pm 0.18 $ & $    101.6\pm 59.0 $ & 4,6,17       \\ 
                   & $      4.3_{-  0.8}^{+  1.3} $ & $     4395\pm  362 $ & $    7.21_{-0.12}^{+0.13} $ & $     -1.47_{-  0.52}^{+  0.41} $ & $    42.38\pm 0.18 $ & $    40.73\pm 0.18 $ & $    115.4\pm 67.6 $ & 18$^b$       \\ 
                   & $\bm{  4.0_{-  0.7}^{+  0.8}}$ & $\bm{ 5142\pm  572}$ & $\bm{7.26_{-0.09}^{+0.09}}$ & $\bm{ -1.10_{-  0.64}^{+  0.60}}$ & $\bm{42.62\pm 0.37}$ & $\bm{40.95\pm 0.33}$ & $\bm{108.3\pm 45.7}$ &              \\ 
      IRAS F12397  & $      9.7_{-  1.8}^{+  5.5} $ & $     1802\pm  560 $ & $    6.79_{-0.45}^{+0.27} $ & $      2.26_{-  0.62}^{+  0.98} $ & $    44.23\pm 0.05 $ & $    42.26\pm 0.04 $ & $     54.2\pm  8.4 $ & 1,2,3$^c$    \\ 
         NGC 4748  & $      5.5_{-  2.2}^{+  1.6} $ & $     1947\pm   66 $ & $    6.61_{-0.23}^{+0.11} $ & $      0.10_{-  0.44}^{+  0.61} $ & $    42.56\pm 0.12 $ & $    40.98\pm 0.10 $ & $    136.8\pm 50.1 $ & 4,14         \\ 
      PG 1307+085  & $    105.6_{- 46.6}^{+ 36.0} $ & $     5059\pm  133 $ & $    8.72_{-0.26}^{+0.13} $ & $     -0.68_{-  0.28}^{+  0.53} $ & $    44.85\pm 0.02 $ & $    43.13\pm 0.06 $ & $     98.4\pm 15.1 $ & 4,5,6,9      \\ 
         NGC 5273  & $      2.2_{-  1.6}^{+  1.2} $ & $     5688\pm  163 $ & $    7.14_{-0.56}^{+0.19} $ & $     -2.50_{-  0.67}^{+  1.33} $ & $    41.54\pm 0.16 $ & $    39.74\pm 0.11 $ & $     82.2\pm 37.1 $ & 19           \\ 
          Mrk 279  & $     16.7_{-  3.9}^{+  3.9} $ & $     5354\pm   32 $ & $    7.97_{-0.12}^{+0.09} $ & $     -0.89_{-  0.30}^{+  0.33} $ & $    43.71\pm 0.07 $ & $    42.12\pm 0.06 $ & $    132.2\pm 28.7 $ & 4,5,6,20     \\ 
      PG 1411+442  & $    124.3_{- 61.7}^{+ 61.0} $ & $     2801\pm   43 $ & $    8.28_{-0.30}^{+0.17} $ & $     -0.23_{-  0.38}^{+  0.63} $ & $    44.56\pm 0.02 $ & $    42.85\pm 0.03 $ & $     99.7\pm  8.2 $ & 4,5,6,9      \\ 
         NGC 5548  & $     19.7_{-  1.5}^{+  1.5} $ & $     4674\pm   63 $ & $    7.92_{-0.04}^{+0.03} $ & $     -1.62_{-  0.49}^{+  0.46} $ & $    43.39\pm 0.10 $ & $    41.79\pm 0.10 $ & $    128.1\pm 40.3 $ & 4,5,6,21     \\ 
                   & $     18.6_{-  2.3}^{+  2.1} $ & $     5418\pm  107 $ & $    8.03_{-0.06}^{+0.05} $ & $     -1.99_{-  0.51}^{+  0.47} $ & $    43.14\pm 0.11 $ & $    41.61\pm 0.13 $ & $    151.3\pm 57.9 $ & 4,5,6,21     \\ 
                   & $     15.9_{-  2.5}^{+  2.9} $ & $     5236\pm   87 $ & $    7.93_{-0.08}^{+0.07} $ & $     -1.68_{-  0.49}^{+  0.45} $ & $    43.35\pm 0.09 $ & $    41.72\pm 0.10 $ & $    119.7\pm 37.9 $ & 4,5,6,21     \\ 
                   & $     11.0_{-  2.0}^{+  1.9} $ & $     5986\pm   95 $ & $    7.89_{-0.09}^{+0.07} $ & $     -2.10_{-  0.52}^{+  0.47} $ & $    43.07\pm 0.11 $ & $    41.52\pm 0.17 $ & $    144.5\pm 66.7 $ & 4,5,6,21     \\ 
                   & $     13.0_{-  1.4}^{+  1.6} $ & $     5930\pm   42 $ & $    7.95_{-0.05}^{+0.05} $ & $     -1.72_{-  0.49}^{+  0.46} $ & $    43.32\pm 0.10 $ & $    41.75\pm 0.09 $ & $    135.9\pm 41.2 $ & 4,5,6,21     \\ 
                   & $     13.4_{-  4.3}^{+  3.8} $ & $     7378\pm   39 $ & $    8.15_{-0.17}^{+0.11} $ & $     -1.64_{-  0.49}^{+  0.45} $ & $    43.38\pm 0.09 $ & $    41.73\pm 0.10 $ & $    114.4\pm 37.0 $ & 4,5,6,21     \\ 
                   & $     21.7_{-  2.6}^{+  2.6} $ & $     6946\pm   79 $ & $    8.31_{-0.06}^{+0.05} $ & $     -1.43_{-  0.48}^{+  0.45} $ & $    43.52\pm 0.09 $ & $    41.82\pm 0.09 $ & $    102.4\pm 30.2 $ & 4,5,6,21     \\ 
                   & $     16.4_{-  1.1}^{+  1.2} $ & $     6623\pm   93 $ & $    8.15_{-0.03}^{+0.03} $ & $     -1.56_{-  0.48}^{+  0.45} $ & $    43.43\pm 0.09 $ & $    41.75\pm 0.10 $ & $    106.3\pm 33.3 $ & 4,5,6,21     \\ 
                   & $     17.5_{-  1.6}^{+  2.0} $ & $     6298\pm   65 $ & $    8.13_{-0.04}^{+0.05} $ & $     -1.85_{-  0.49}^{+  0.46} $ & $    43.24\pm 0.10 $ & $    41.72\pm 0.10 $ & $    153.5\pm 50.7 $ & 4,5,6,21     \\ 
                   & $     26.5_{-  2.2}^{+  4.3} $ & $     6177\pm   36 $ & $    8.30_{-0.04}^{+0.07} $ & $     -1.33_{-  0.48}^{+  0.45} $ & $    43.59\pm 0.09 $ & $    41.87\pm 0.10 $ & $     98.1\pm 30.0 $ & 4,5,6,21     \\ 
                   & $     24.8_{-  3.0}^{+  3.2} $ & $     6247\pm   57 $ & $    8.28_{-0.06}^{+0.05} $ & $     -1.45_{-  0.48}^{+  0.45} $ & $    43.51\pm 0.09 $ & $    41.83\pm 0.09 $ & $    106.6\pm 31.5 $ & 4,5,6,21     \\ 
                   & $      6.5_{-  3.7}^{+  5.7} $ & $     6240\pm   77 $ & $    7.69_{-0.37}^{+0.27} $ & $     -2.04_{-  0.51}^{+  0.47} $ & $    43.11\pm 0.11 $ & $    41.64\pm 0.13 $ & $    172.8\pm 66.1 $ & 4,5,6,21     \\ 
                   & $     14.3_{-  7.3}^{+  5.9} $ & $     6478\pm  108 $ & $    8.07_{-0.31}^{+0.15} $ & $     -2.03_{-  0.51}^{+  0.47} $ & $    43.11\pm 0.11 $ & $    41.55\pm 0.14 $ & $    139.8\pm 55.7 $ & 4,5,6,21     \\ 
                   & $      6.3_{-  2.3}^{+  2.6} $ & $     6396\pm  167 $ & $    7.70_{-0.20}^{+0.15} $ & $     -2.27_{-  0.55}^{+  0.49} $ & $    42.96\pm 0.13 $ & $    41.12\pm 0.10 $ & $     74.4\pm 27.3 $ & 4,22         \\ 
                   & $      4.2_{-  1.3}^{+  0.9} $ & $    12771\pm   71 $ & $    8.12_{-0.16}^{+0.08} $ & $     -2.19_{-  0.51}^{+  0.47} $ & $    43.01\pm 0.11 $ & $    41.33\pm 0.10 $ & $    105.3\pm 35.4 $ & 4,14         \\ 
                   & $     12.4_{-  3.9}^{+  2.7} $ & $    11481\pm  574 $ & $    8.50_{-0.17}^{+0.09} $ & $     -2.21_{-  0.53}^{+  0.48} $ & $    42.99\pm 0.11 $ & $    41.27\pm 0.10 $ & $     96.6\pm 34.4 $ & 4,13$^a$     \\ 
                   & $\bm{ 17.6_{-  4.7}^{+  6.4}}$ & $\bm{ 7241\pm 2200}$ & $\bm{8.10_{-0.16}^{+0.16}}$ & $\bm{ -1.80_{-  0.33}^{+  0.33}}$ & $\bm{43.29\pm 0.20}$ & $\bm{41.64\pm 0.23}$ & $\bm{117.2\pm 25.9}$ &              \\ 
      PG 1426+015  & $     95.0_{- 37.1}^{+ 29.9} $ & $     7113\pm  160 $ & $    8.97_{-0.22}^{+0.12} $ & $     -1.51_{-  0.28}^{+  0.47} $ & $    44.63\pm 0.02 $ & $    42.83\pm 0.04 $ & $     80.1\pm  9.2 $ & 4,5,6,9      \\ 
          Mrk 817  & $     19.0_{-  3.7}^{+  3.9} $ & $     4711\pm   49 $ & $    7.92_{-0.09}^{+0.08} $ & $     -0.81_{-  0.35}^{+  0.35} $ & $    43.79\pm 0.05 $ & $    42.07\pm 0.05 $ & $     98.0\pm 16.5 $ & 4,5,6,7      \\ 
                   & $     15.3_{-  3.5}^{+  3.7} $ & $     5237\pm   67 $ & $    7.91_{-0.11}^{+0.09} $ & $     -0.98_{-  0.35}^{+  0.35} $ & $    43.67\pm 0.05 $ & $    42.00\pm 0.06 $ & $    108.5\pm 20.4 $ & 4,5,6,7      \\ 
                   & $     33.6_{-  7.6}^{+  6.5} $ & $     4767\pm   72 $ & $    8.17_{-0.11}^{+0.08} $ & $     -0.98_{-  0.35}^{+  0.35} $ & $    43.67\pm 0.05 $ & $    41.92\pm 0.05 $ & $     91.1\pm 15.3 $ & 4,5,6,7      \\ 
                   & $     14.0_{-  3.5}^{+  3.4} $ & $     5627\pm   30 $ & $    7.94_{-0.12}^{+0.09} $ & $     -0.73_{-  0.35}^{+  0.35} $ & $    43.84\pm 0.05 $ & $    41.77\pm 0.05 $ & $     43.2\pm  7.1 $ & 4,13         \\ 
                   & $\bm{ 19.9_{-  6.7}^{+  9.9}}$ & $\bm{ 5348\pm  536}$ & $\bm{7.99_{-0.14}^{+0.14}}$ & $\bm{ -0.87_{-  0.22}^{+  0.22}}$ & $\bm{43.74\pm 0.09}$ & $\bm{41.93\pm 0.14}$ & $\bm{ 78.5\pm 34.3}$ &              \\ 
         Mrk 1511  & $      5.7_{-  0.8}^{+  0.9} $ & $     4171\pm  137 $ & $    7.29_{-0.07}^{+0.07} $ & $     -0.34_{-  0.24}^{+  0.24} $ & $    43.16\pm 0.06 $ & $    41.52\pm 0.06 $ & $    115.5\pm 23.1 $ & 18$^b$       \\ 
          Mrk 290  & $      8.7_{-  1.0}^{+  1.2} $ & $     4543\pm  227 $ & $    7.55_{-0.07}^{+0.07} $ & $     -0.85_{-  0.23}^{+  0.23} $ & $    43.17\pm 0.06 $ & $    41.64\pm 0.06 $ & $    153.0\pm 29.0 $ & 4,13$^a$     \\ 
          Mrk 486  & $     23.7_{-  2.7}^{+  7.5} $ & $     1942\pm   67 $ & $    7.24_{-0.06}^{+0.12} $ & $      0.55_{-  0.32}^{+  0.20} $ & $    43.69\pm 0.05 $ & $    42.12\pm 0.04 $ & $    135.9\pm 20.3 $ & 1,2,3        \\ 
          Mrk 493  & $     11.6_{-  2.6}^{+  1.2} $ & $      778\pm   12 $ & $    6.14_{-0.11}^{+0.04} $ & $      1.88_{-  0.21}^{+  0.33} $ & $    43.11\pm 0.08 $ & $    41.35\pm 0.05 $ & $     87.4\pm 18.1 $ & 1,2,3        \\ 
      PG 1613+658  & $     40.1_{- 15.2}^{+ 15.0} $ & $     9074\pm  103 $ & $    8.81_{-0.21}^{+0.14} $ & $     -0.97_{-  0.31}^{+  0.45} $ & $    44.77\pm 0.02 $ & $    43.00\pm 0.03 $ & $     86.7\pm  7.6 $ & 4,5,6,9      \\ 
      PG 1617+175  & $     71.5_{- 33.7}^{+ 29.6} $ & $     6641\pm  190 $ & $    8.79_{-0.28}^{+0.15} $ & $     -1.50_{-  0.33}^{+  0.58} $ & $    44.39\pm 0.02 $ & $    42.74\pm 0.05 $ & $    114.8\pm 15.1 $ & 4,5,6,9      \\ 
      PG 1700+518  & $    251.8_{- 38.8}^{+ 45.9} $ & $     2252\pm   85 $ & $    8.40_{-0.08}^{+0.08} $ & $      1.08_{-  0.17}^{+  0.17} $ & $    45.59\pm 0.01 $ & $    43.78\pm 0.02 $ & $     78.9\pm  4.5 $ & 4,5,6,9      \\ 
         3C 390.3  & $     23.6_{-  6.7}^{+  6.2} $ & $    12694\pm   13 $ & $    8.87_{-0.15}^{+0.10} $ & $     -3.35_{-  0.65}^{+  0.60} $ & $    43.68\pm 0.10 $ & $    42.29\pm 0.05 $ & $    206.2\pm 50.7 $ & 4,5,6,23     \\ 
                   & $     46.4_{-  3.2}^{+  3.6} $ & $    13211\pm   28 $ & $    9.20_{-0.03}^{+0.03} $ & $     -2.12_{-  0.51}^{+  0.51} $ & $    44.50\pm 0.03 $ & $    42.78\pm 0.04 $ & $     97.1\pm 10.0 $ & 4,24         \\ 
                   & $\bm{ 44.5_{- 17.0}^{+ 27.6}}$ & $\bm{12796\pm  361}$ & $\bm{9.18_{-0.23}^{+0.23}}$ & $\bm{ -2.62_{-  0.96}^{+  0.95}}$ & $\bm{44.43\pm 0.58}$ & $\bm{42.60\pm 0.35}$ & $\bm{108.8\pm 58.8}$ &              \\ 
     KA 1858+4850  & $     13.5_{-  2.3}^{+  2.0} $ & $     1820\pm   79 $ & $    6.94_{-0.09}^{+0.07} $ & $      0.75_{-  0.21}^{+  0.25} $ & $    43.43\pm 0.05 $ & $    41.89\pm 0.04 $ & $    146.9\pm 21.1 $ & 25$^d$       \\ 
         NGC 6814  & $      6.6_{-  0.9}^{+  0.9} $ & $     3323\pm    7 $ & $    7.16_{-0.06}^{+0.05} $ & $     -1.64_{-  0.80}^{+  0.46} $ & $    42.12\pm 0.28 $ & $    40.50\pm 0.28 $ & $    121.6\pm112.2 $ & 4,14         \\ 
          Mrk 509  & $     79.6_{-  5.4}^{+  6.1} $ & $     3015\pm    2 $ & $    8.15_{-0.03}^{+0.03} $ & $     -0.52_{-  0.14}^{+  0.13} $ & $    44.19\pm 0.05 $ & $    42.61\pm 0.04 $ & $    132.7\pm 19.1 $ & 4,5,6,7      \\ 
      PG 2130+099  & $      9.6_{-  1.2}^{+  1.2} $ & $     2450\pm  188 $ & $    7.05_{-0.10}^{+0.08} $ & $      1.69_{-  0.20}^{+  0.23} $ & $    44.20\pm 0.03 $ & $    42.65\pm 0.03 $ & $    142.1\pm 13.6 $ & 4,8$^a$      \\ 
         NGC 7469$^e$ & $     10.8_{-  1.3}^{+  3.4} $ & $     4369\pm    6 $ & $    7.60_{-0.06}^{+0.12} $ & $      0.90_{-  1.87}^{+  1.83} $ & $    43.51\pm 0.11 $ & $    41.60\pm 0.10 $ & $     63.0\pm 21.0 $ & 4,26         \\ 
                   & $      4.5_{-  0.8}^{+  0.7} $ & $     1722\pm   30 $ & $    6.42_{-0.09}^{+0.06} $ & $      0.63_{-  1.90}^{+  1.85} $ & $    43.32\pm 0.12 $ & $    41.71\pm 0.09 $ & $ 124.9\pm 44.3 $ & 27,5         \\ 
                   & $\bm{  6.5_{-  3.0}^{+  5.8}}$ & $\bm{ 4343\pm 2859}$ & $\bm{6.92_{-0.84}^{+0.84}}$ & $\bm{  0.76_{-  1.35}^{+  1.31}}$ & $\bm{43.43\pm 0.15}$ & $\bm{41.66\pm 0.10}$ & $\bm{ 86.9\pm 47.1}$ &              \\ 
\enddata
\tablecomments{Ref:
  1. Paper I; 
  2. Paper II; 
  3. Paper III; 
  4. Bentz et al. (2013); 
  5. Collin et al. (2006); 
  6. Kaspi et al. (2005); 
  7. Peterson et al. (1998); 
  8. Grier et al. (2012); 
  9. Kaspi et al. (2000); 
  10. Santos-Lleo et al. (1997); 
  11. This paper; 
  12. Bentz et al. (2009a); 
  13. Denney et al. (2010); 
  14. Bentz et al. (2009b); 
  15. Stirpe et al. (1994); 
  16. Bentz et al. (2006); 
  17. Denney et al. (2006); 
  18. Barth et al. (2013); 
  19. Bentz et al. (2014); 
  20. Santos-Lleo et al. (2001); 
  21. Peterson et al. (2002) and references therein; 
  22. Bentz et al. (2007); 
  23. Dietrich et al. (1998); 
  24. Dietrich et al. (2012); 
  25. Pei et al. (2014); 
  26. Peterson et al. (2014); 
  27. Collier et al. (1998).\\
~~~Superscript $a$ means that the literature does not provide FWHMs from mean spectra or H$\beta$ 
fluxes with narrow-line components subtracted. For these, we scanned and digitized the mean 
spectra from the papers, then calculated those numbers ourselves. $b$ means that the AGN 
continuum is obtained through spectral fitting decomposition (Ref. 18). $c$ for IRAS F12397, 
we use fluxes for the case with local absorption correction (see the details in Papers I and 
III). $d$ means that the contribution of host galaxy is provided in Ref. 25. $e$ the
virial products of the two independent campaigns are quite different, yielding a large
error bars on the mean $\bhm$ and $\mathdotM$. In Appendix B, NGC 7469 is shown as an SEAMBH
from Collier et al. (1988) and Peterson et al. (2014). Both campaigns have high quality data. 
Numbers in boldface are the weighted averages of all the measurements of this object.
The details of the average scheme are given in the main text.
}
\end{deluxetable}

\clearpage
\begin{figure}[t!]
\begin{center}
\includegraphics[angle=0,width=0.7\textwidth]{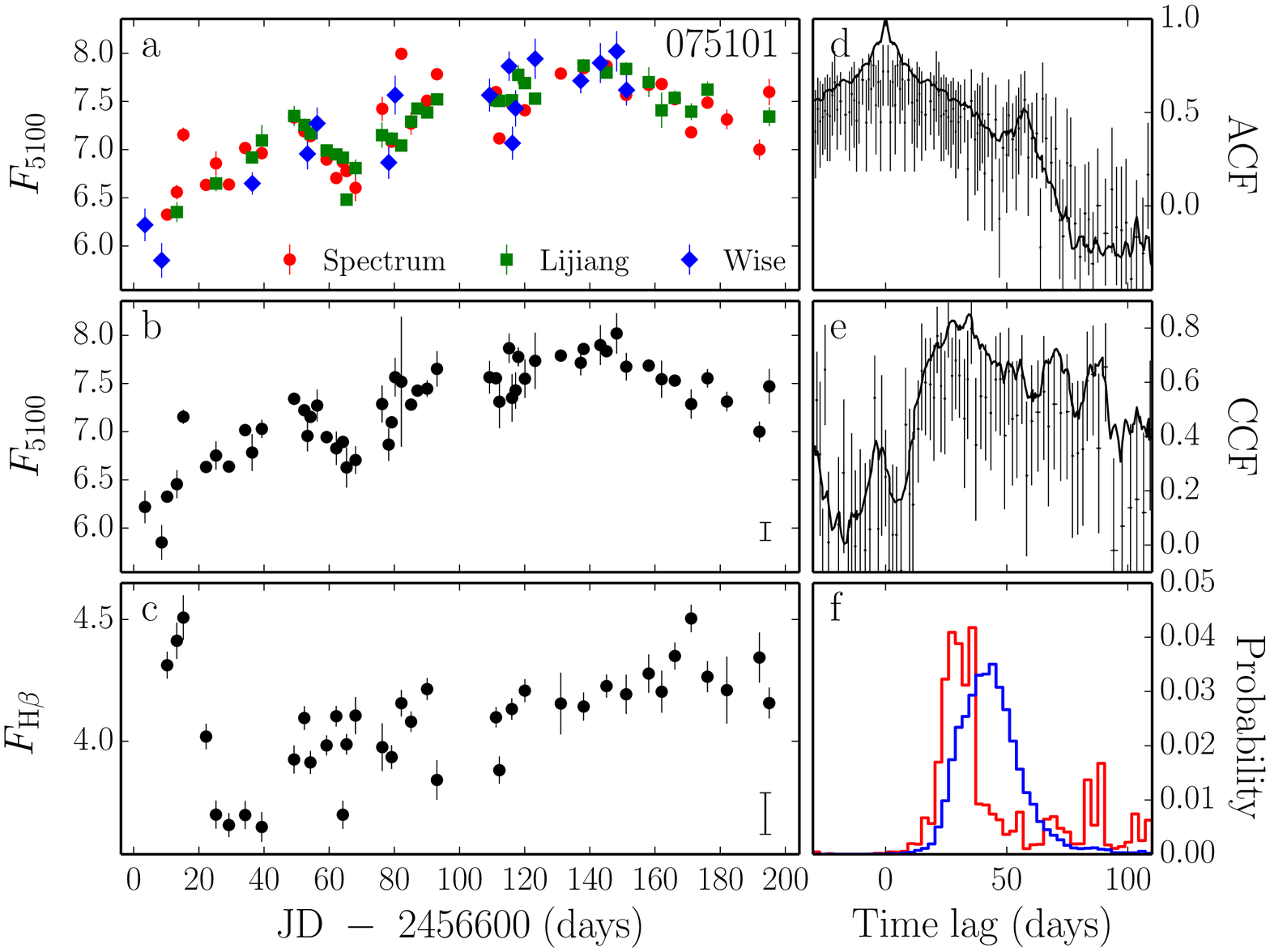}
\includegraphics[angle=0,width=0.7\textwidth]{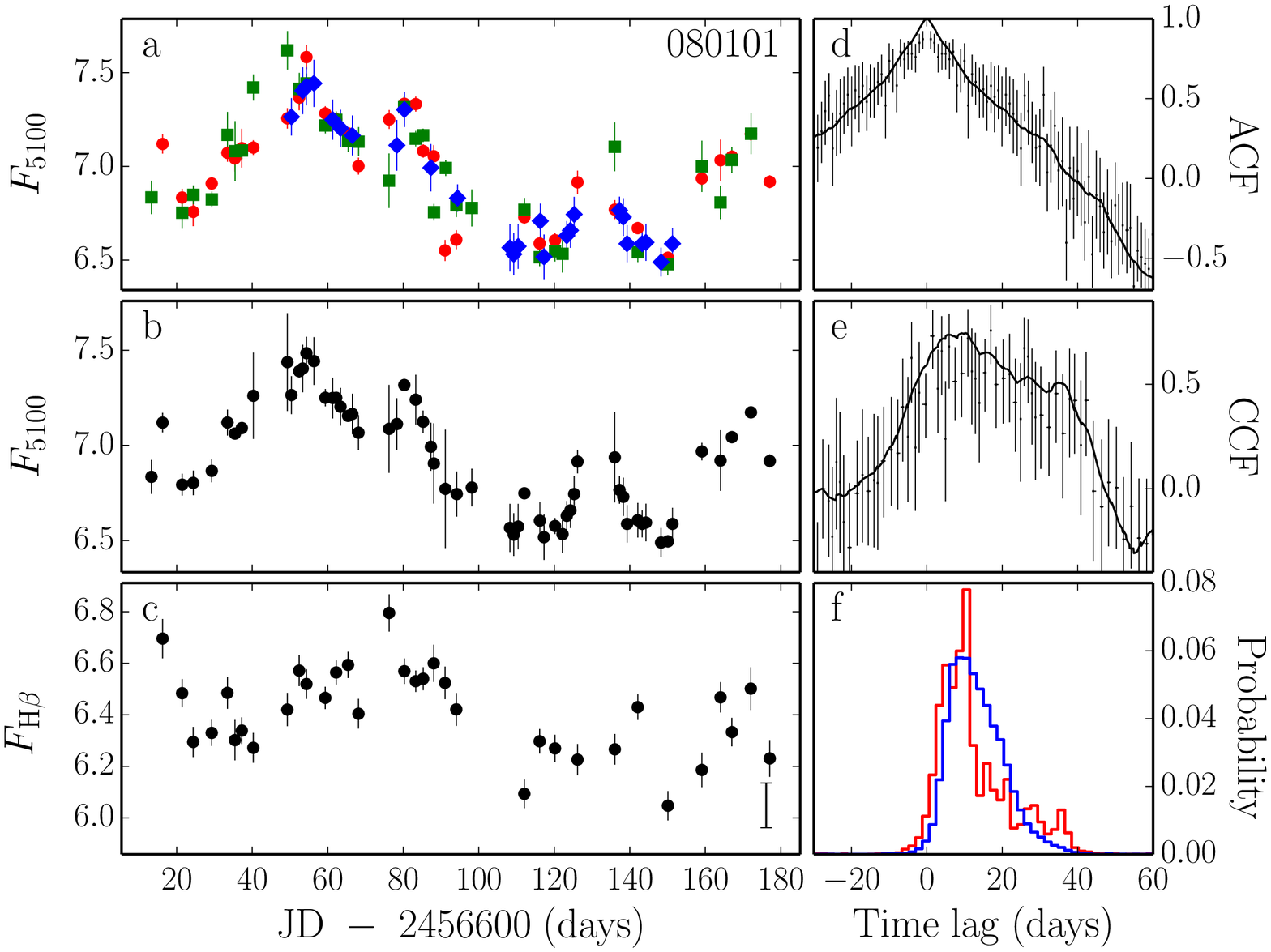}
\end{center}
\vglue -0.5cm
\caption{\footnotesize Light curves and cross correlation results. Each object has six panels: 
({\it a, b, c}) are light curves of the inter-calibrated continuum, combined 5100\AA\, 
continuum, and H$\beta$ emissions, respectively; ({\it d, e, f}) are auto correlation function 
(ACF) of the combined continuum, cross correlation function (CCF) of the 
combined continuum and H$\beta$ emission and the Monte-Carlo simulations of peaks (red) and 
centroid (blue) of lags, respectively. In panels {\it d} and {\it e}, the solid lines 
show the results of ICCF method and the  points with error bars are from ZDCF ($Z-$transformed 
discrete correlation function). $F_{5100}$ and $F_{\rm H\beta}$ are in unit of 
$10^{-16}{\rm erg~s^{-1}~cm^{-2}\AA^{-1}}$ and $10^{-14}{\rm erg~s^{-1}~cm^{-2}}$ for all quasars, 
respectively. In panel ({\it a}) of J075101, the red dots, green squares and blue diamonds are 
fluxes from Lijiang spectroscopy, photometry and Wise photometry, respectively. The colour points 
in other objects have the same meanings with that of J075101.
Bars with terminals as systematic errors are plotted in the corners of the panels
(see Paper I for details). 
}
\end{figure}

\begin{figure*}[t!]
\begin{center}
\includegraphics[angle=0,width=0.7\textwidth]{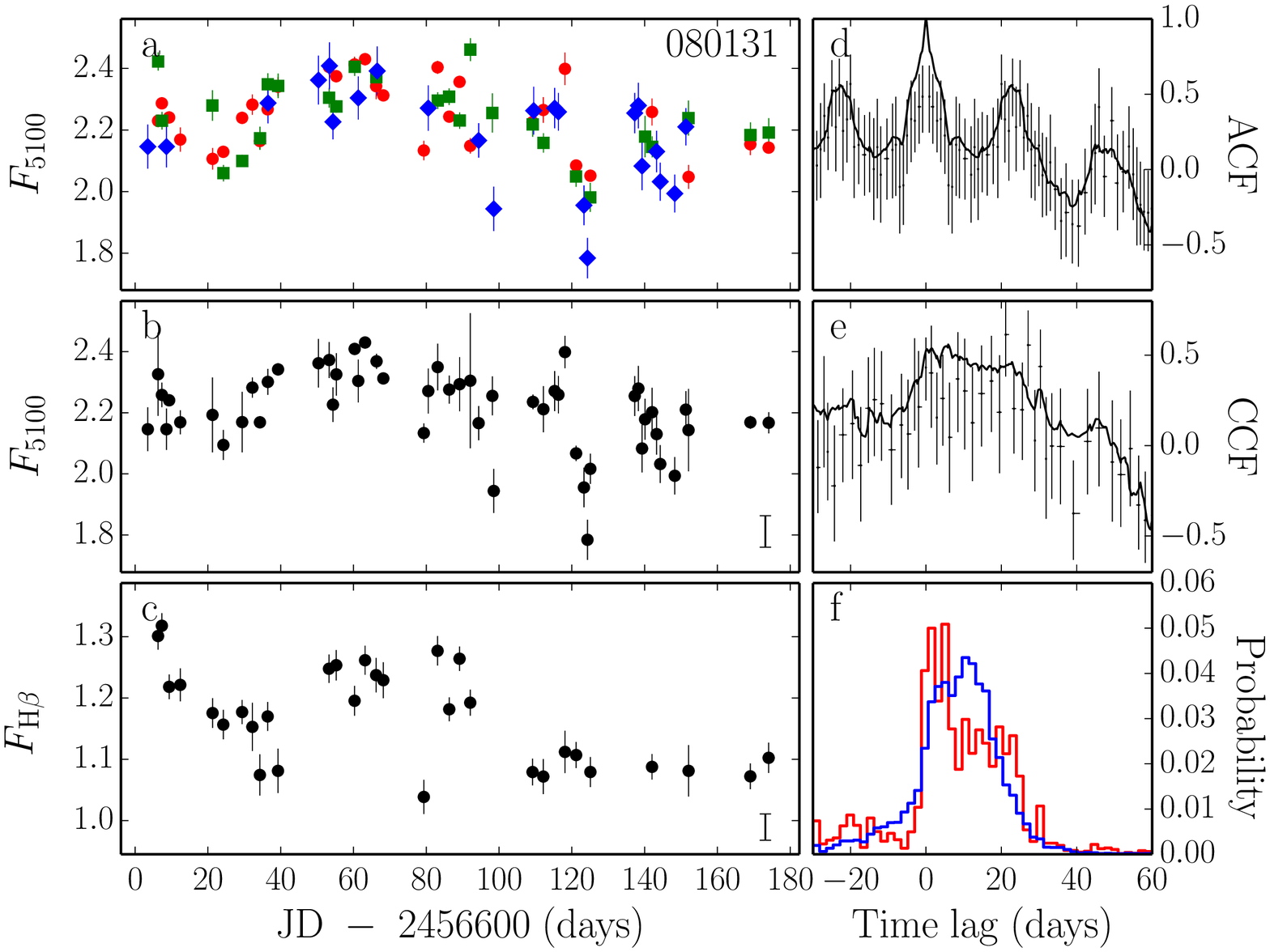}
\includegraphics[angle=0,width=0.7\textwidth]{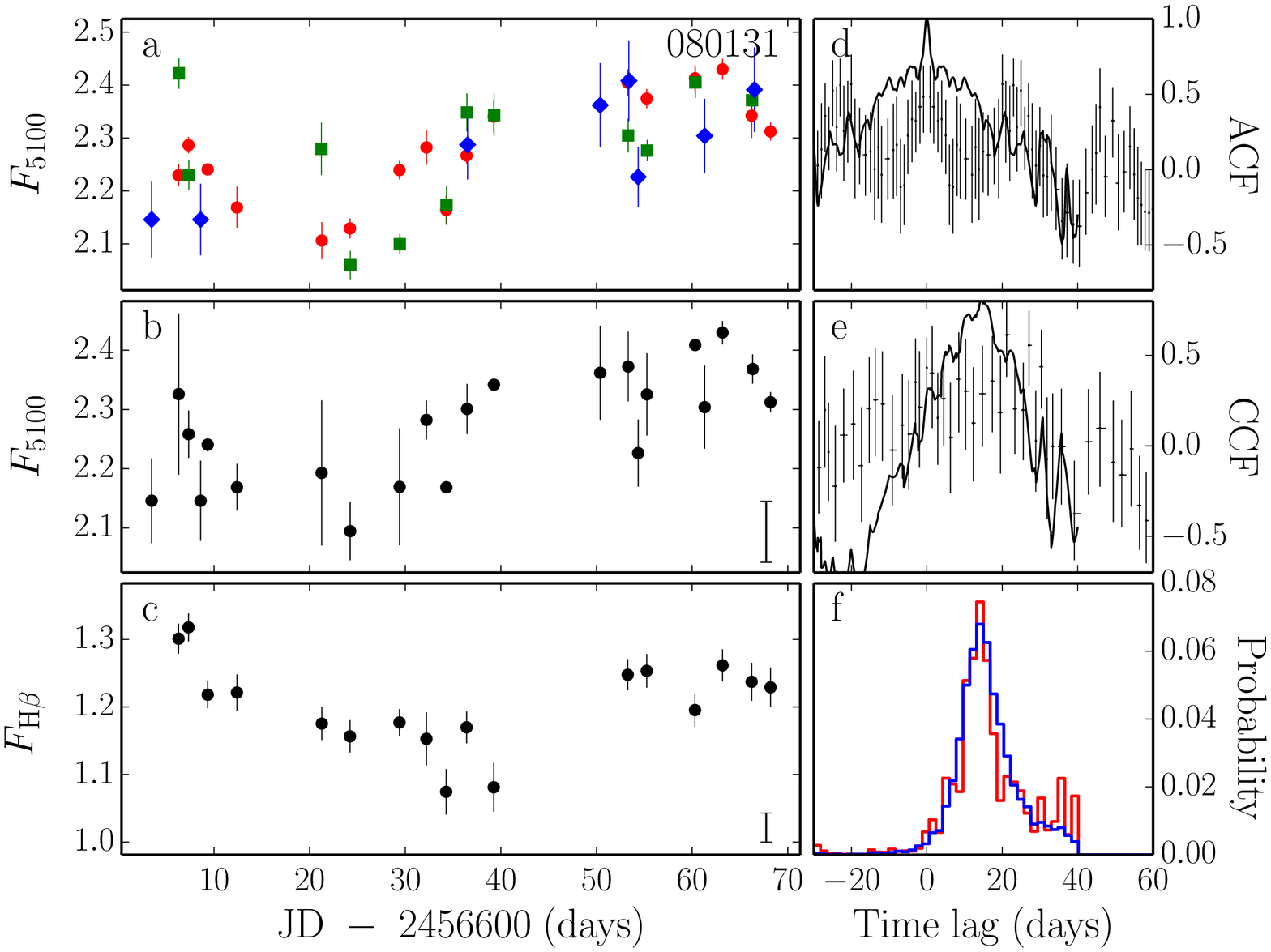}
\end{center}
{Figure 1 {\it continued.}}
\end{figure*}

\begin{figure*}[t!]
\begin{center}
\includegraphics[angle=0,width=0.7\textwidth]{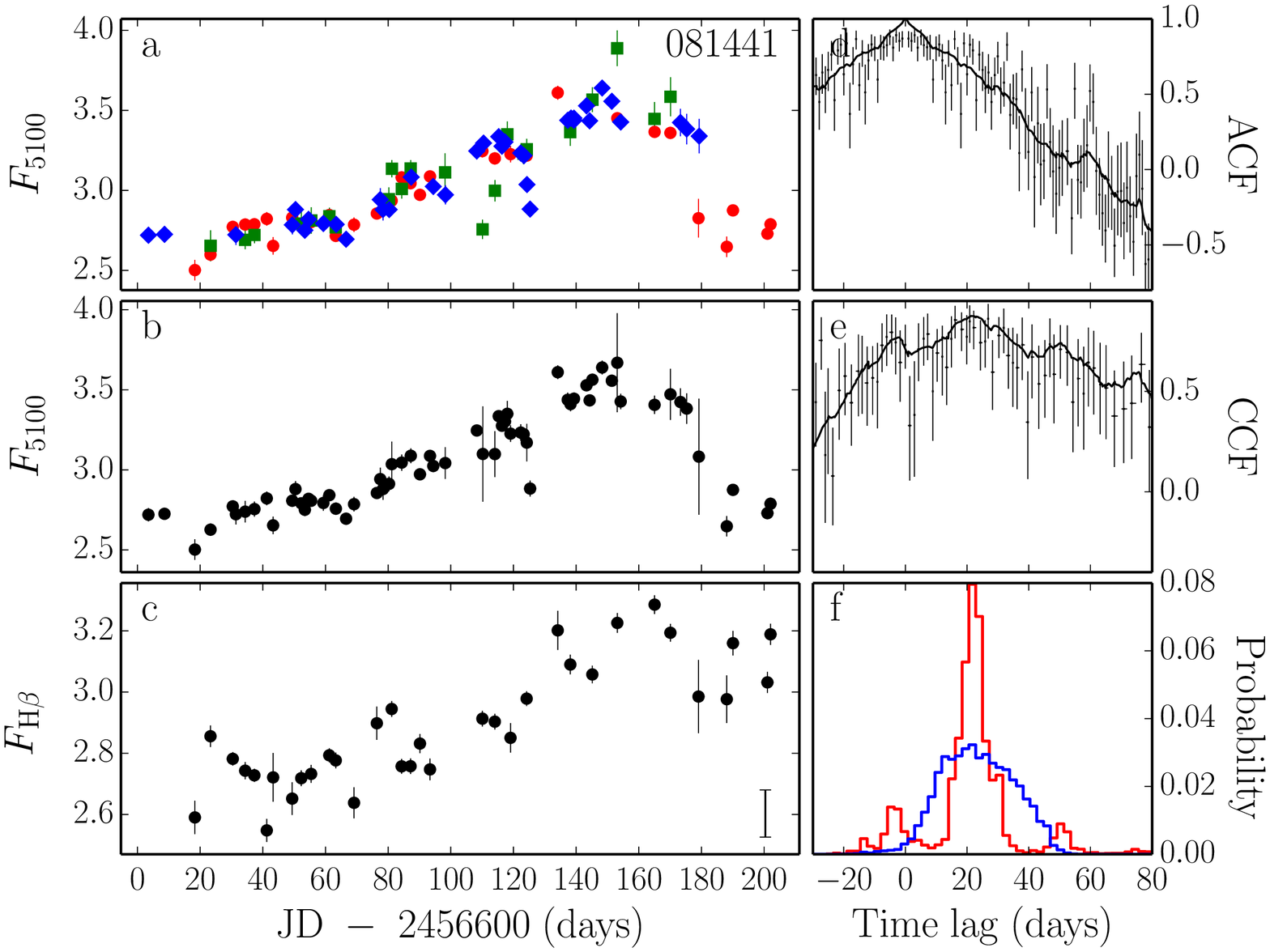}
\includegraphics[angle=0,width=0.7\textwidth]{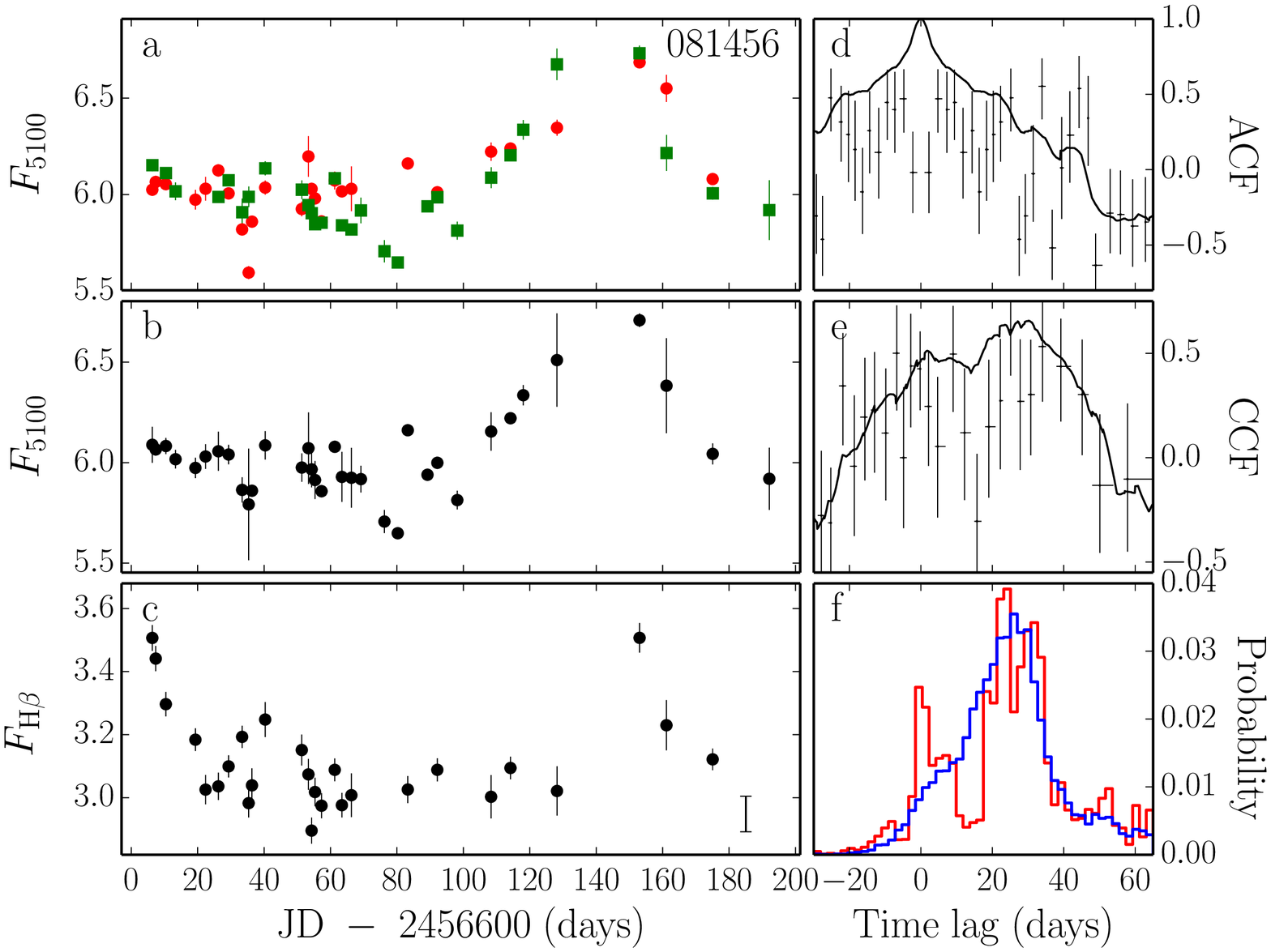}
\end{center}
{Figure 1 {\it continued.}}
\end{figure*}

\begin{figure*}[t!]
\begin{center}
\includegraphics[angle=0,width=0.7\textwidth]{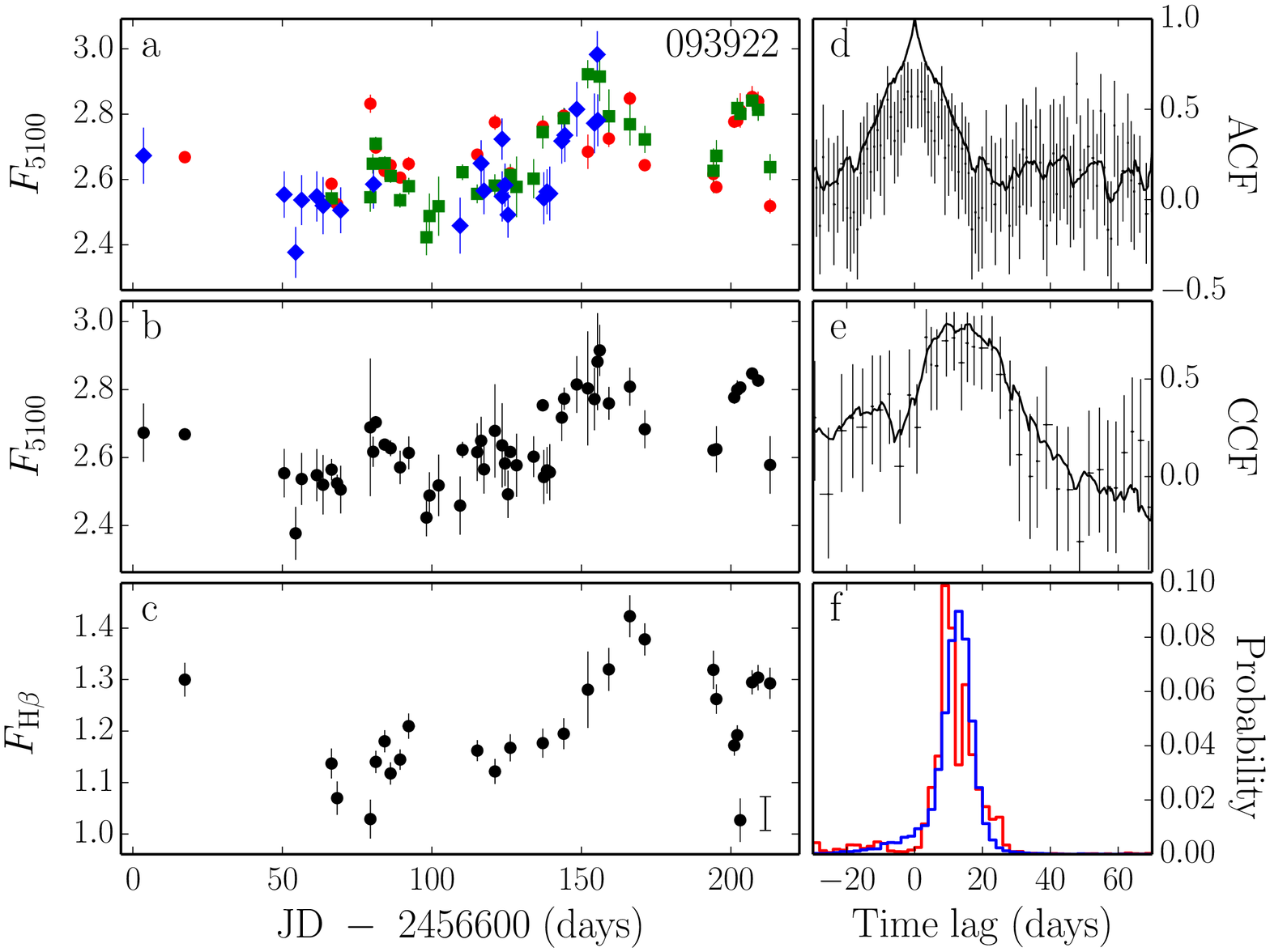}
\includegraphics[angle=0,width=0.7\textwidth]{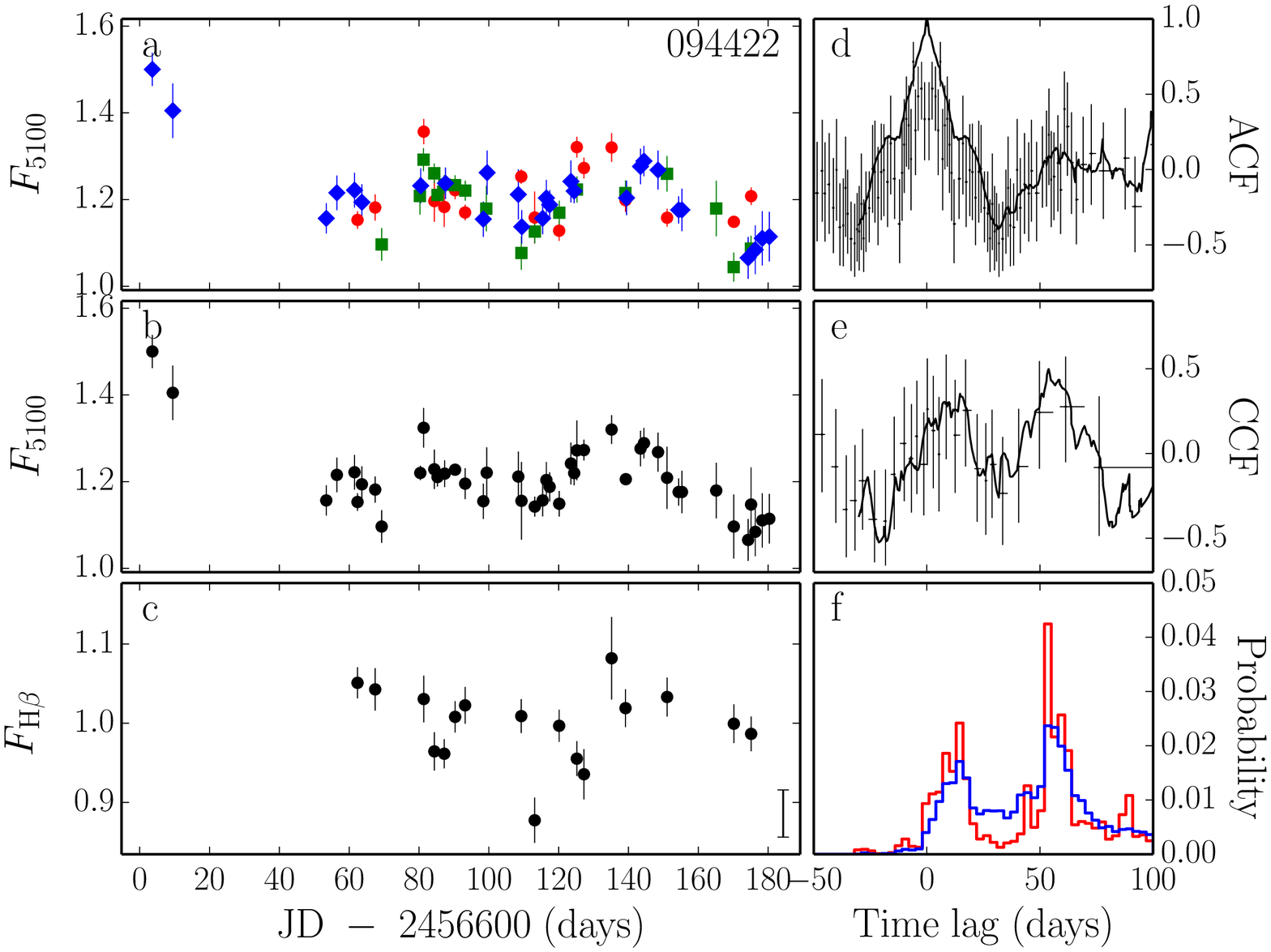}
\end{center}
{Figure 1 {\it continued.}}
\end{figure*}

\begin{figure*}[t!]
\begin{center}
\includegraphics[angle=0,width=0.7\textwidth]{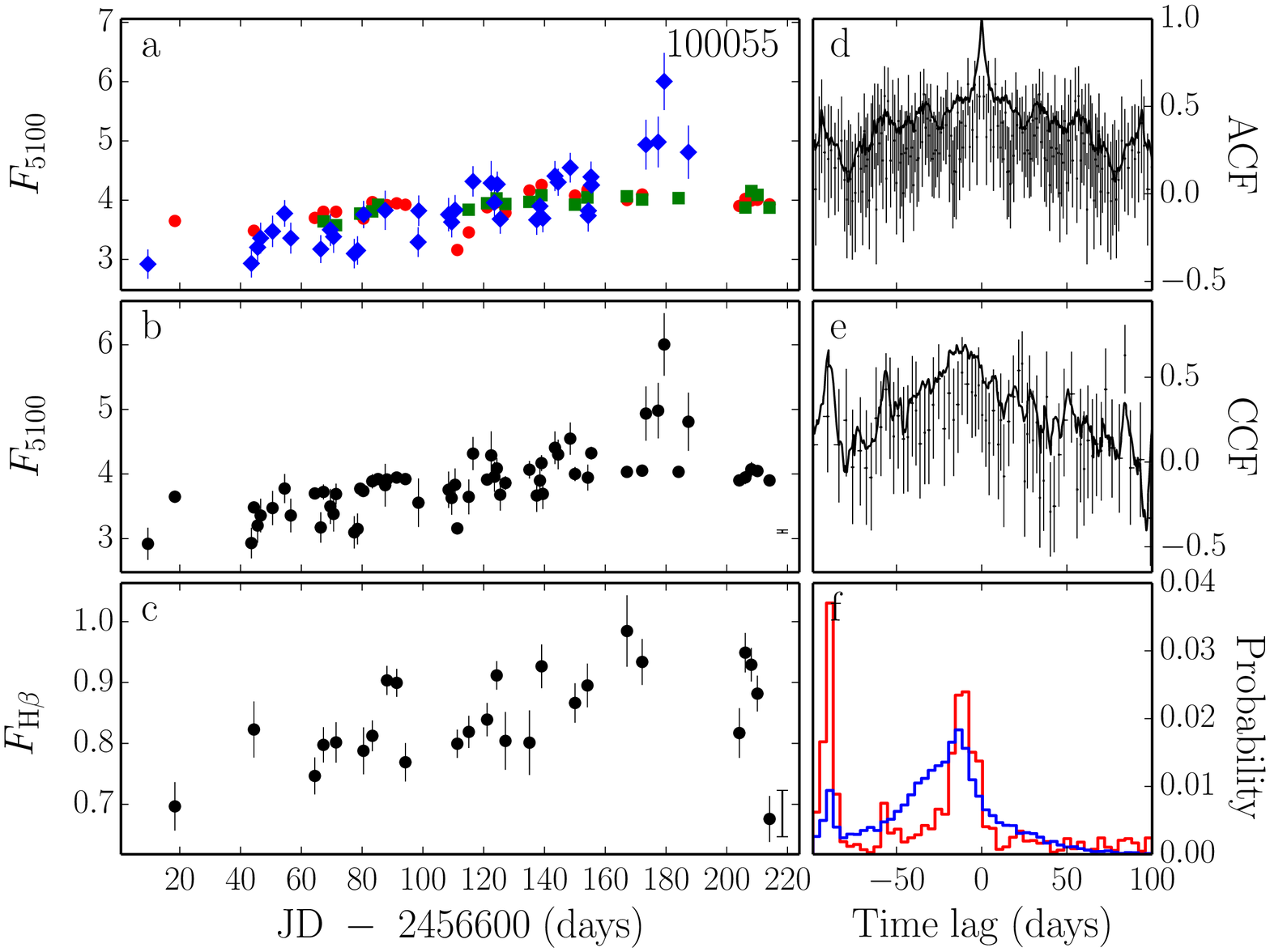}
\end{center}
{Figure 1 {\it continued.}}
\end{figure*}

\clearpage
\begin{figure*}[t!]
\begin{center}
\includegraphics[angle=0,width=0.9\textwidth]{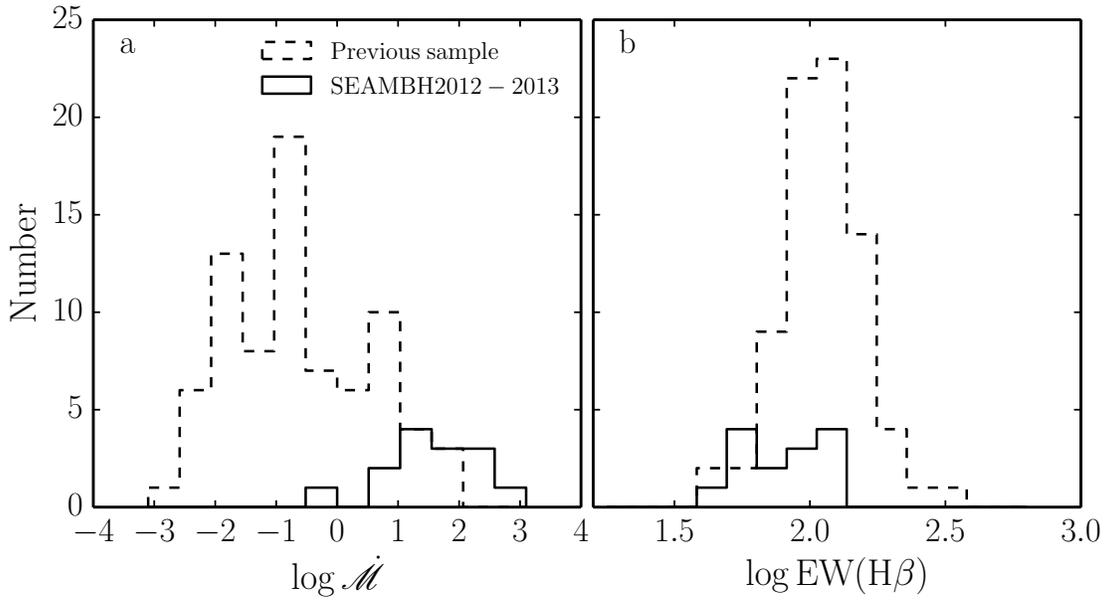}
\end{center}
\vglue -0.5cm
\caption{\footnotesize 
The distribution of dimensionless accretion rates of the mapped AGNs and quasars. The 
previous sample refers to the sample of all mapped AGNs summarized by Bentz et al. 
(2013) (the repeated monitored AGNs are regarded as individual ones), and adds four 
recently mapped AGNs: NGC 7469 updated by Peterson et al. (2014), KA 1858+4850 (Pei 
et al. 2014), Mrk 1511 (Barth et al. 2013) and NGC 5273 (Bentz et al. 2014).  The 
dashed line is for the previous sample and the solid is for the SEAMBH sample. Our 
campaign selected those candidate sources with extremely high accretion rates 
($\mathdotM\gtrsim 10$). Though some sources monitored previously have 
$\mathdotM\sim 10$, most of them have lower accretion rates. {\it Right}
panel shows the distribution of EW(H$\beta$), showing that SEAMBH sample tends to 
have low EW(H$\beta$).
}
\label{dotm}
\end{figure*}

\clearpage
\begin{figure*}[t!]
\begin{center}
\includegraphics[angle=0,width=1.0\textwidth]{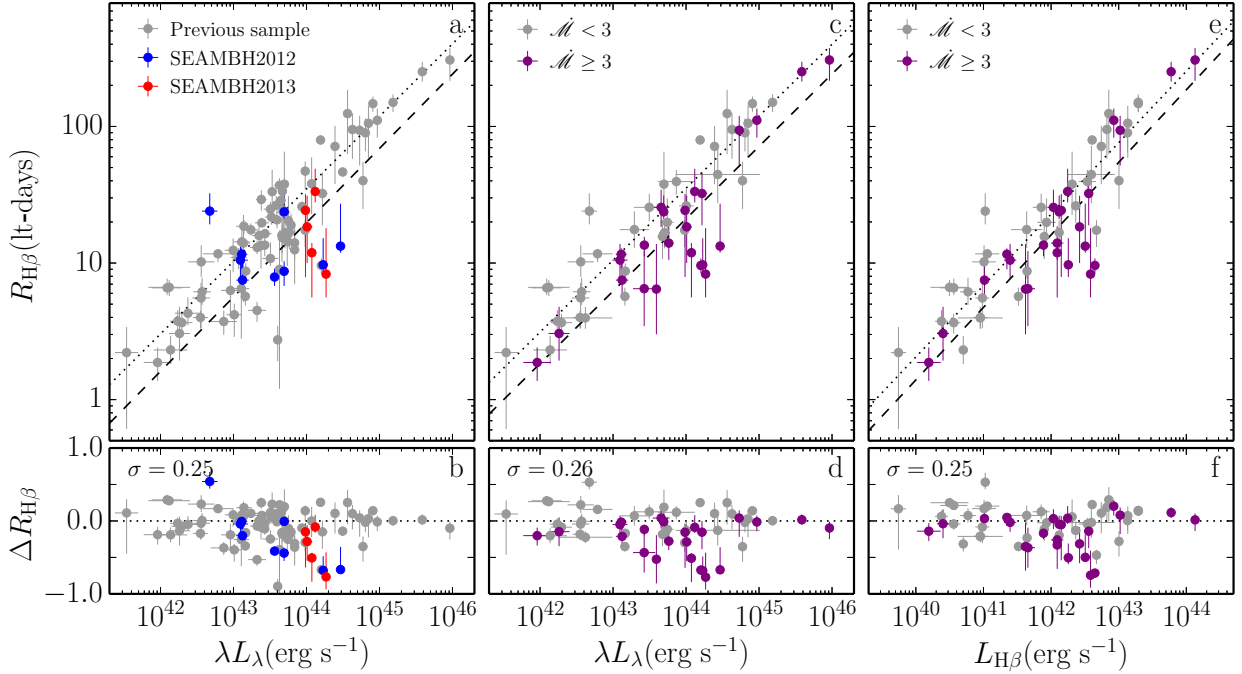}
\end{center}
\vglue -0.5cm
\caption{\footnotesize 
The $\rhb-L_{5100}$ and $\rhb-\Lhb$ relationships (panel {\it a, c, e}) 
and the deviations of $\rblr$ from the regression relationships (panel {\it b, d, f}). 
The dotted is the fit for the $\mathdotM<3$ sample whereas the dashed
is the fit for the $\mathdotM>3$ sample in all panels.
Panels ({\it a, b}) show the comparison of the previous sample and the SEAMBH 
campaigns. Clearly SEAMBHs are below the $\rblrl$ relation, and the scatter of the relation
increases with the inclusion of the SEAMBH2012/2013 samples (see the intrinsic scatter 
given by the numbers below Equations 4 and 5). Panel {\it b} shows 
$\Delta R_{\rm H\beta}=\log \left(\rblr/\prblr\right)$, where $\prblr$ 
is given by Equation (4b). 
Panels ({\it c,d}) show the two samples of $\mathdotM\ge 3$ and $\mathdotM<3$ AGNs with averaged
RM values for the $\rhb-L_{5100}$ relation and the residual of $\Delta \rhb$, where $\prblr$ is 
given by Equation 5b. 
Panels ({\it e, f}) show the $\rhb-\Lhb$ and $\Delta\rhb-\mathdotM$ relations. 
}
\label{r-l}
\end{figure*}

\clearpage
\begin{figure*}[t!]
\begin{center}
\includegraphics[angle=0,width=0.95\textwidth]{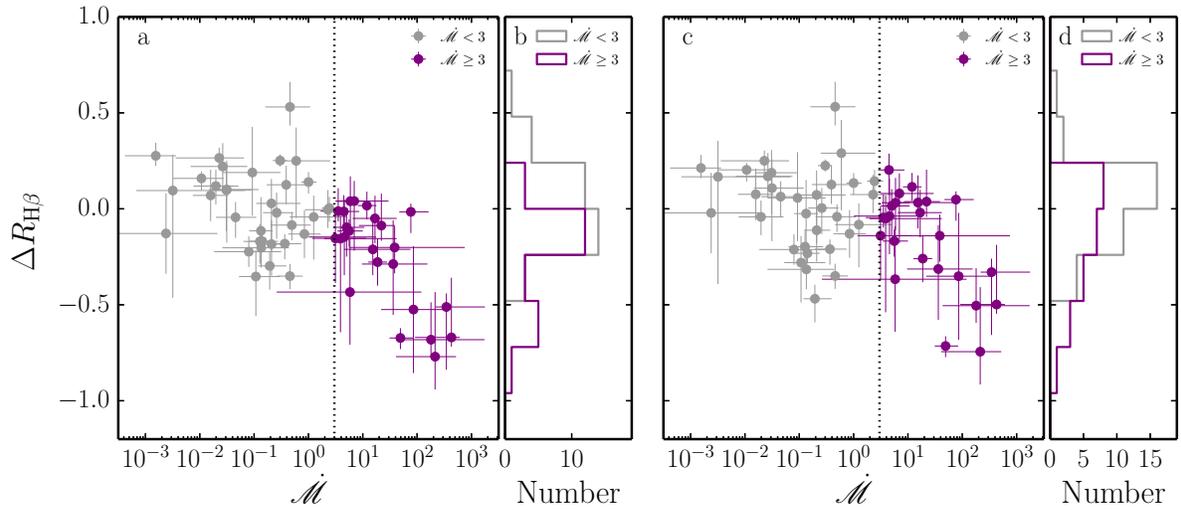}
\end{center}
\vglue -0.7cm 
\caption{\footnotesize 
Panels ({\it a, b}) and ({\it c, d}) show the $\Delta\rhb-\mathdotM$ relation and $\Delta\rhb$ 
distributions for $\rhb-L_{5100}$ and $\rhb-\Lhb$ relations, respectively. 
The dotted lines indicate $\mathdotM=3$
in panel {\it a} and {\it c}. 
Both panels show that $\Delta \rhb \sim 0$ up to 
$\mathdotM\sim 3 $, beyond which $|\Delta \rhb|$ increases with $\mathdotM$.  This implies that, 
relative to normal AGNs of the same 5100\AA\, luminosity, SEAMBHs have H$\beta$ lags that become 
systematically shorter with increasing accretion rate.
}
\label{r-l}
\end{figure*}

\clearpage
\begin{figure*}[t!]
\begin{center}
\includegraphics[angle=0,width=0.95\textwidth]{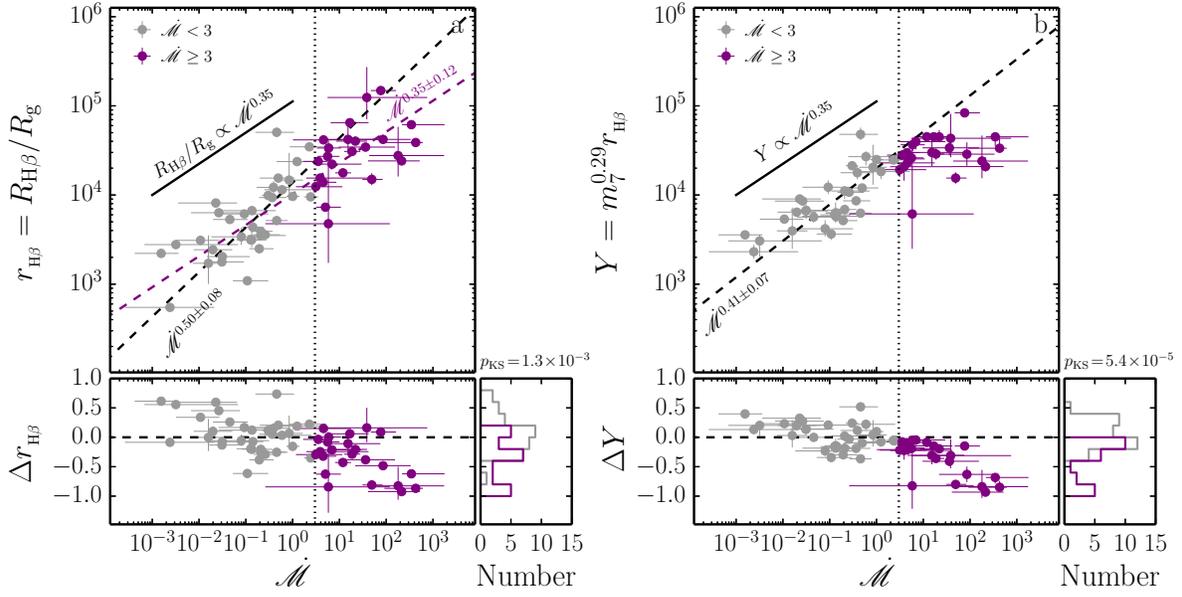}
\end{center}
\vglue -0.7cm
\caption{\footnotesize 
Black hole mass-scaled BLR and the radius-mass parameter versus dimensionless accretion rate.
In each panel, there are three plots to show their relation and deviations. Correlations and 
probabilities of KS tests are marked in the plots. 
Panel {\it a} shows that the $\mathdotM<3$ sample has a steeper correlation than that given 
by Equation (9). 
Panel {\it b} shows that
SEAMBHs deviate from this relation and that 
$\Ythin$ is independent of $\mathdotM$. Clearly $Y$ behaves in a different way in the two accretion 
rate groups.}
\label{r-l}
\end{figure*}

\clearpage

\begin{figure*}[t!]
\begin{center}
\includegraphics[angle=0,width=0.5\textwidth]{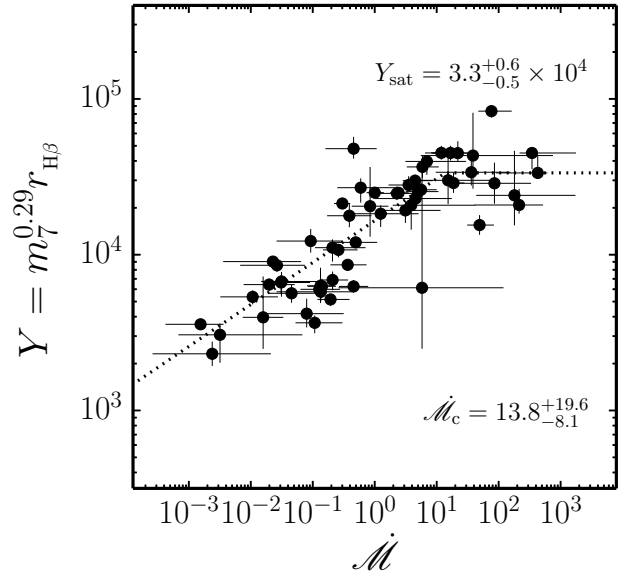}
\end{center}
\vglue -0.7cm
\caption{\footnotesize 
Determination of the saturated$-Y$ and the transition accretion rate. We find 
$\Ythin\propto \mathdotM^{0.27\pm0.04}$, supporting the SS73 disk model below $\mathdotM_c$
and saturation beyond the numbers listed inside the panel supporting the idea of 
slim accretion disks in SEAMBHs.
The outlier with large error bars is NGC 7469 (see Table 7). } 
\label{r-l}
\end{figure*}

\clearpage

\begin{figure*}[t!]
\begin{center}
\includegraphics[angle=0,width=0.95\textwidth]{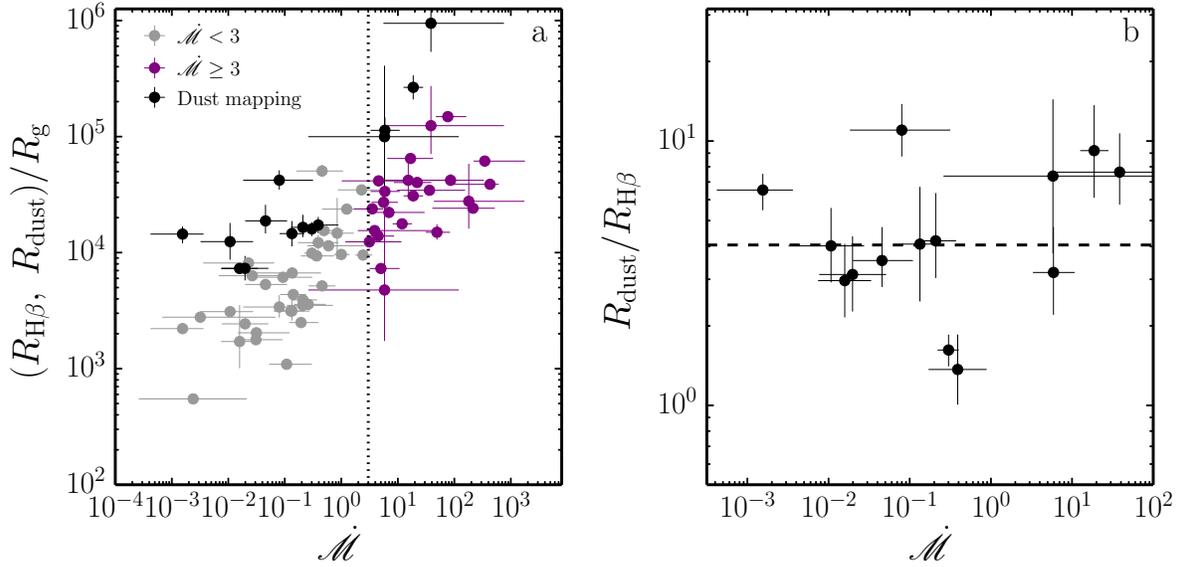}
\end{center}
\vglue -0.7cm
\caption{\footnotesize Black hole mass-scaled BLR (purple and grey points) and 
dust reverberation radii (black points). Panel {\it a} 
shows $R_{\rm dust}/\Rg$ and $\rblr/\Rg$ versus $\mathdotM$. 
Panel {\it b} shows ratios of $R_{\rm dust}/\rblr$ as a function of $\mathdotM$. Note the 
large scatter in both low and high $\mathdotM$ groups.
}
\label{r-l}
\end{figure*}

\clearpage
\appendix

\section{Mean and RMS spectra}
To supplement the material presented in the body of the paper, we provide in Figure 8 the 
mean and RMS spectra of the objects in the SEAMBH2013 sample calculated by the standard method 
(e.g., Peterson et al. 2004). Prior to the calculation of these spectra, we corrected the data 
for the slight wavelength shift caused by the relatively wide aperture used in 2013
($5^{\prime\prime}$) using the \oiii$\lambda5007$ line as our wavelength reference. 
We define
\begin{equation}
\overline{F}(\lambda)=\frac{1}{N}\sum_{i=1}^NF_i(\lambda) \, ,
\end{equation}
and
\begin{equation}
S(\lambda)=\left\{\frac{1}{N}\sum_{i=1}^N\left[F_i(\lambda)-\overline{F}(\lambda)\right]^2\right\}^{1/2},
\end{equation}
where $F_i(\lambda)$ is the $i-$th spectrum and $N$ the total number of spectra obtained during the campaign. 
The line dispersion is 
\begin{equation}
\sigma_{\rm line}^2(\lambda)=\langle\lambda^2\rangle-\lambda_0^2,
\end{equation}
where
$
\lambda_0=\int\lambda P(\lambda)d\lambda/\int P(\lambda) d\lambda,
$
$
\langle\lambda^2\rangle=\int\lambda^2 P(\lambda)d\lambda/\int P(\lambda) d\lambda,
$
and $P(\lambda)$ is the line profile. The calculated values of $\sigma_{\rm line}$ 
are listed in Table 6. 

\section{$\rblr$ Correlations using the direct method}
The correlations listed in the main body of the paper were calculated using the average scheme 
where each object is represented by one point. The results shown in this section 
are for the direct scheme, where each observing campaign is represented 
by one point. 

The following equations correspond to Equation (6)
\begin{subequations}
\begin{empheq}[left={\log \left(\rhb/{\rm ltd}\right)=\empheqlbrace}]{align}
&(1.29\pm 0.02)+(0.52\pm0.03)\log L_{\rm H\beta,42}&({\rm entire~ sample}),\\
&(1.35\pm0.03)+(0.53\pm0.03)\log L_{\rm H\beta,42}&(\mathdotM<3),\\
&(1.17\pm0.05)+(0.55\pm0.06)\log L_{\rm H\beta,42}&(\mathdotM\ge3),
\end{empheq}
\end{subequations}
with scatters of $\sigma_{\rm in}=(0.20,0.15,0.24)$ for (B1a,B1b,B1c), respectively.
For $\Delta \rhb-\mathdotM$ relation, we have
\begin{equation}
\Delta \rhb=(0.49\pm0.14)-(0.58\pm0.10)\log \mathdotM ~~~ ({\rm for ~\mathdotM\ge3} ),
\end{equation}
and 
\begin{equation}
\Delta \rhb=(0.44\pm0.15)-(0.49\pm0.11)\log \mathdotM ~~~({\rm for~\mathdotM\ge3}),
\end{equation}
with $\sigma_{\rm in}=(0.01,0.11)$ for deviations from $\rblrl$ and $\rhb-\Lhb$ relations, 
respectively. These correlations are shown in Figures 9 and 10.

For the $Y-\mathdotM$ relation as shown by Figure 11, we obtained
\begin{equation}
\mathdotM_c=10.8_{-5.7}^{+12.0},~~\Ysat=3.0_{-0.4}^{+0.4}\times 10^4,
~~{\rm and}~~k_0=0.28\pm 0.03,
\end{equation} 
from Figure 12, yielding the critical radius of
\begin{equation}
\rrhb=3.0^{+0.4}_{-0.4}\times 10^4~m_7^{0.29};~~{\rm or}~~
R_{\rm crit}({\rm H\beta})=17.1^{+2.3}_{-2.3} ~m_7^{1.29}~{\rm ltd}.
\end{equation}
The results are very similar, and entirely consistent with the results of the average 
scheme presented in the main body of the paper.

\clearpage
\begin{figure*}[t!]
\begin{center}
\includegraphics[angle=0,width=0.95\textwidth]{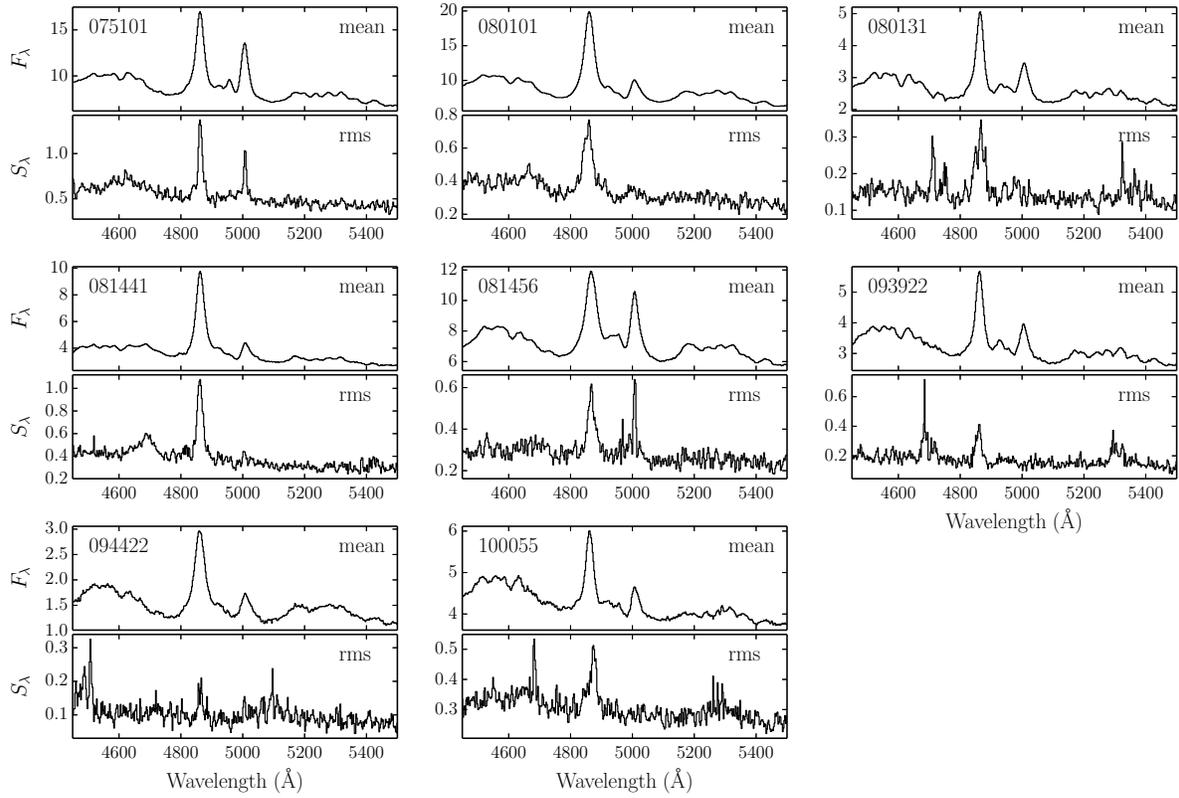}
\end{center}
\vglue -0.7cm
\caption{\footnotesize Mean and RMS spectra (observed flux vs. rest-frame wavelength) 
of the objects in the SEAMBH2013 group.
}
\label{r-l}
\end{figure*}

\clearpage
\begin{figure*}[t!]
\begin{center}
\includegraphics[angle=0,width=0.9\textwidth]{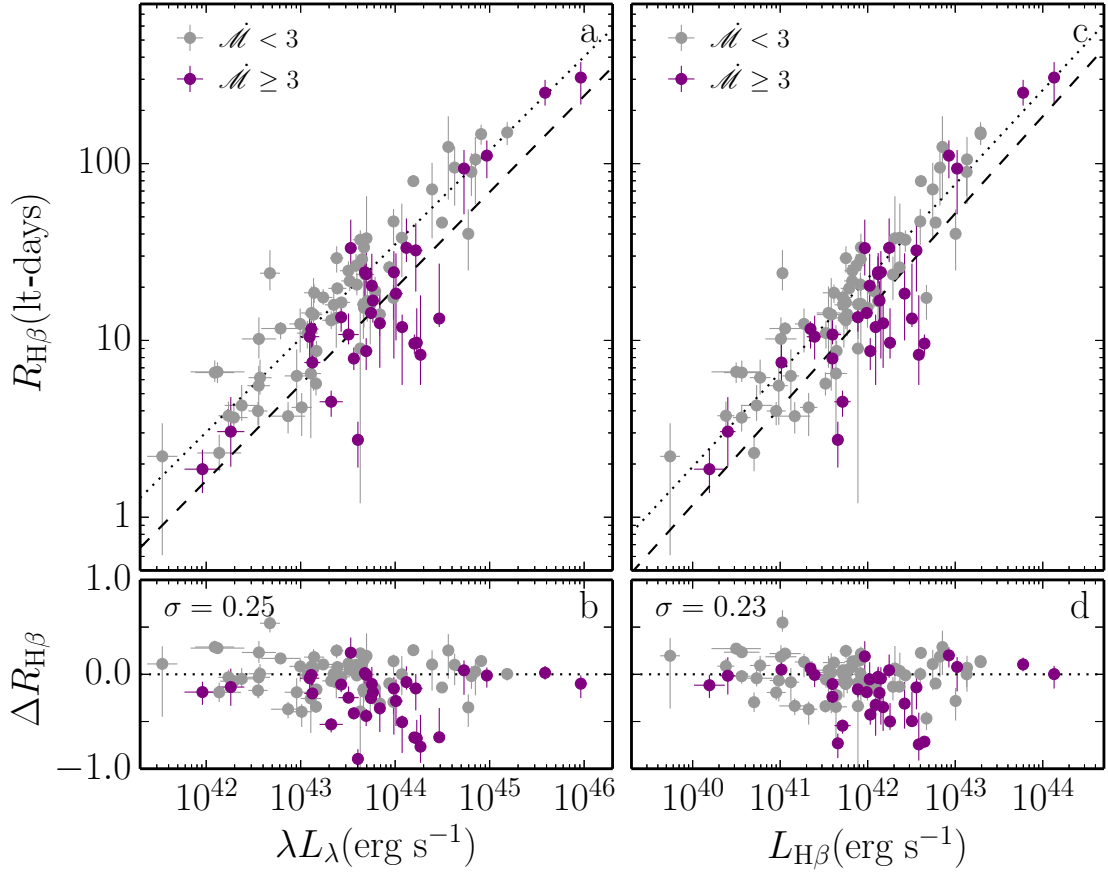}
\end{center}
\vglue -0.7cm
\caption{\footnotesize Same as Figure 3 but for the direct scheme. The {\it left} panel and Figure 3{\it a}
are identical, the {\it right} panel corresponds to Figure 3{\it e}.
}
\label{r-l}
\end{figure*}

\clearpage
\begin{figure*}[t!]
\begin{center}
\includegraphics[angle=0,width=0.95\textwidth]{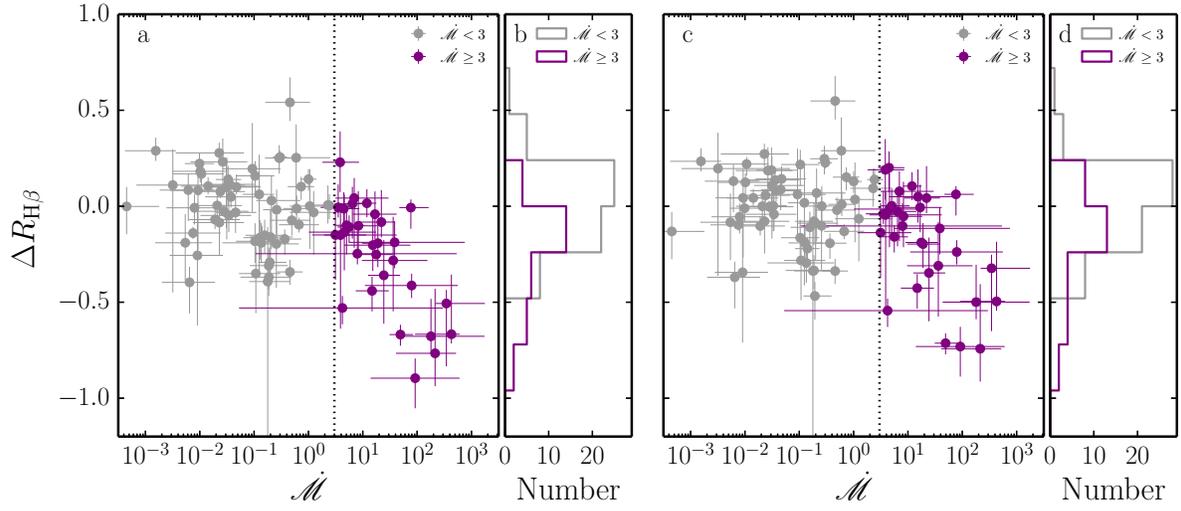}
\end{center}
\vglue -0.7cm
\caption{\footnotesize Same as Figure 4 but for the direct scheme. The object with the largest error bar
is NGC 7469.
}
\label{r-l}
\end{figure*}

\clearpage
\begin{figure*}[t!]
\begin{center}
\includegraphics[angle=0,width=0.95\textwidth]{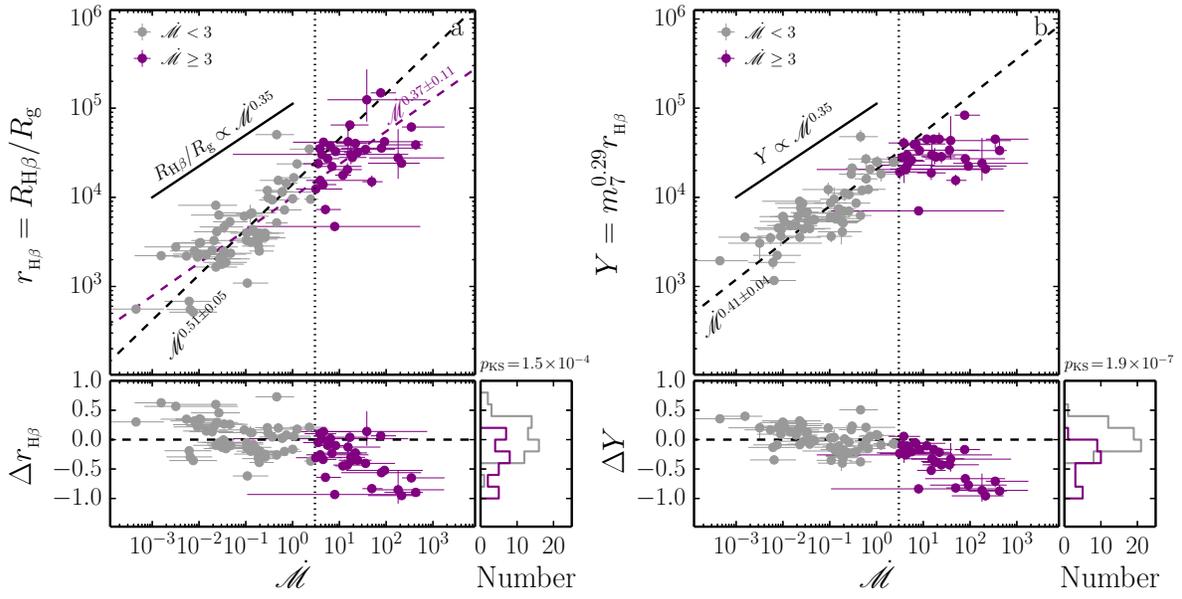}
\end{center}
\vglue -0.7cm
\caption{\footnotesize
Same as Figure 5 but for the direct scheme. }
\label{r-l}
\end{figure*}

\clearpage
\begin{figure*}[t!]
\begin{center}
\includegraphics[angle=0,width=0.5\textwidth]{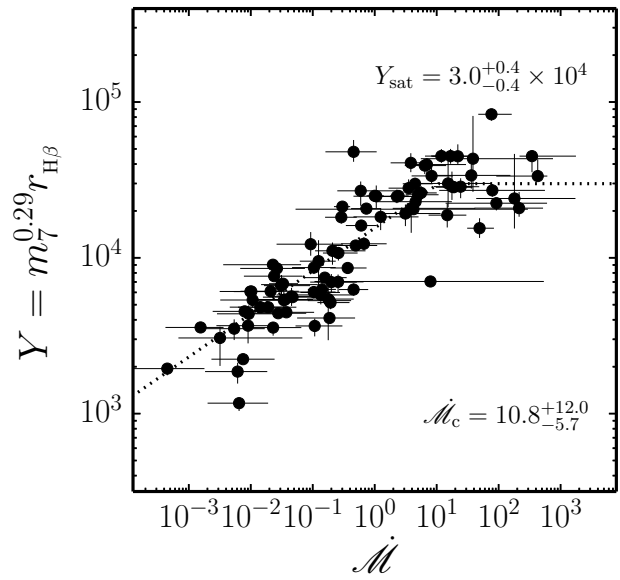}
\end{center}
\vglue -0.7cm
\caption{\footnotesize Same as Figure 6 for the direct scheme.}
\label{r-l}
\end{figure*}

\end{document}